\documentclass[12pt,a4paper,titlepage]{article}
\usepackage[left=2.00cm, right=2.00cm, top=2.00cm, bottom=2.00cm]{geometry}
\usepackage[utf8]{inputenc}
\usepackage[T1]{fontenc}
\usepackage{amsmath}
\usepackage{amsfonts}
\usepackage{amssymb}
\usepackage{graphicx}
\usepackage[table]{xcolor}
\usepackage{booktabs}
\usepackage{threeparttable}
\usepackage{array}
\usepackage{booktabs}
\usepackage{float}
\usepackage{tabularx}
\usepackage{hyperref}
\usepackage{physics}
\usepackage[bottom]{footmisc}
\usepackage{tikz}
\usetikzlibrary{quantikz2}
\usepackage[]{mdframed}
\usepackage{algpseudocode}
\usepackage{algorithm}
\usepackage{subcaption}

\usepackage{titling}

\newcolumntype{L}[1]{>{\raggedright\let\newline\\\arraybackslash\hspace{0pt}}m{#1}}
\newcolumntype{C}[1]{>{\centering\let\newline\\\arraybackslash\hspace{0pt}}m{#1}}
\newcolumntype{R}[1]{>{\raggedleft\let\newline\\\arraybackslash\hspace{0pt}}m{#1}}

\usepackage{tikz}
\usepackage{pgfgantt}

\begin{document}


 



\begin{center}
  \Large Feedback-Based Quantum Algorithm for Excited States Calculation 
\end{center}
\begin{center}
   Salahuddin Abdul Rahman$^1$, Özkan Karabacak$^2$, Rafal Wisniewski $^1$ 
\end{center}
\begin{center}
  1 Automation and Control section, Department of electronic systems, Aalborg University, Aalborg, Denmark \\
  2 Department of Mechatronics Engineering, Kadir Has University,  Istanbul, Turkey
\end{center}


\section*{Abstract}
Recently, feedback-based quantum algorithms have been introduced to calculate the ground states of Hamiltonians, inspired by quantum Lyapunov control theory. This paper aims to generalize these algorithms to the problem of calculating an eigenstate of a given Hamiltonian, assuming that the lower energy eigenstates are known. To this aim, we propose a new design methodology that combines the layer-wise construction of the quantum circuit in feedback-based quantum algorithms with a new feedback law based on a new Lyapunov function to assign the quantum circuit parameters. We present two approaches for evaluating the circuit parameters: one based on the expectation and overlap estimation of the terms in the feedback law and another based on the gradient of the Lyapunov function. We demonstrate the algorithm through an illustrative example and through an application in quantum chemistry. To assess its performance, we conduct numerical simulations and execution on IBM's superconducting quantum computer.

\section{Introduction}
\label{sec:paperB:introduction}

Calculating ground states and excited states of Hamiltonians is paramount in various domains, including quantum chemistry, condensed matter physics and combinatorial optimization. Variational quantum algorithms (VQAs) are the leading algorithms for NISQ devices that can approximately calculate ground states and excited states \cite{bharti2022noisy}. These algorithms are tailored to meet the demands of NISQ devices. However, the efficacy of VQAs faces challenges, including the intricate design of ansatz structures and the need to tackle computationally demanding classical optimization problems for parameter updates within parameterized quantum circuits \cite{cerezo2021variational}. Some of the proposals to tackle the challenge of ansatz design are based on a layer-wise ansatz construction, named as adaptive derivative assembled pseudo-Trotter variational quantum eigensolver (ADAPT-VQE) algorithm \cite{grimsley2019adaptive} and adaptive derivative assembled problem tailored quantum approximate optimization algorithm (ADAPT-QAOA)\cite{zhu2022adaptive}. To tackle the challenge of the classical optimization problem,  suitable classical optimizers were proposed \cite{fernandez2022study}.

In \cite{magann2022lyapunov,magann2022feedback},  Magann et al. introduced the feedback-based algorithm for quantum optimization (FALQON) as a novel approach for solving quadratic unconstrained binary optimization (QUBO) problems. FALQON constructs the quantum circuit layer by layer and determines the parameters of the next layer through measurements of qubits from the preceding layer to estimate the gradient of the cost function with respect to the circuit parameters. This methodology circumvents classical optimization and exhibits a monotonic enhancement of approximate solutions with increasing circuit depth. In addition, the feedback-based quantum algorithm (FQA) was proposed in \cite{larsen2024feedback} to generalize FALQON to prepare the ground states of Hamiltonians, specifically for molecular Hamiltonians expressed in the second quantization and for the Fermi-Hubbard model. In \cite{magann2023randomized}, a randomization technique is incorporated to enhance the performance of feedback-based quantum algorithms. FALQON can also be utilized as an initialization technique for the quantum approximate optimization algorithm (QAOA) parameters, potentially increasing its performance \cite{magann2022lyapunov}.

Investigating the excited states of Hamiltonians holds significance across diverse applications, such as quantum chemistry and combinatorial optimization. In quantum chemistry, determining the energy spectrum of a quantum system is crucial, especially when examining drug discovery and catalysis reaction rates. The dynamics of molecules are characterized by their energy spectra; therefore, understanding these spectra is crucial in molecular design \cite{pyrkov2023quantum}. Similarly, in combinatorial optimization, calculating excited states is of paramount importance. For example, in problems such as the shortest vector problem (SVP), the solution is encoded as the first excited state of the emerging Hamiltonian, obtained by transforming SVP translated as a QUBO problem into an Ising Hamiltonian \cite{joseph2021two, ura2023analysis}. In literature, several quantum algorithms have been proposed to find the excited states, such as VQAs \cite{higgott2019variational,colless2018computation}, imaginary-time evolution algorithms \cite{tsuchimochi2023improved,yeter2020practical}, quantum annealing algorithms \cite{ura2023analysis,seki2021excited}, and others \cite{wen2024full,bauman2020toward,cortes2022quantum}.

In \cite{higgott2019variational}, Higgott et al. introduced the variational quantum deflation (VQD) algorithm, which extends the applicability of the Variational Quantum Eigensolver (VQE) to compute excited states. They utilized Hotelling's deflation method \cite{mackey2008deflation} to construct an effective Hamiltonian, which possesses its lowest eigenstate as the first excited state of the original Hamiltonian, and its lowest eigenvalue corresponds to the energy of that particular state. Accordingly, they define a new cost function to minimize the energy of the effective Hamiltonian instead of the original Hamiltonian. In this paper, using a similar methodology and utilizing tools from quantum Lyapunov control (QLC) theory, we propose the feedback-based quantum algorithm for excited states preparation (FQAE). Our contribution is to propose a new feedback law based on a new Lyapunov function to assign the quantum circuit parameters and to introduce efficient hybrid methods to estimate the proposed feedback law. We integrate this new approach to the layer-wise construction of the quantum circuit in feedback-based quantum algorithms to find the excited states of the problem Hamitlonian. We propose a suitable Lyapunov function using a new operator that encodes the $m$'th excited state of the original Hamiltonian as its ground state. We give two different approaches to evaluate the controllers using both quantum and classical computers. In the first approach, we evaluate the controller by reducing it to different terms, including expected values and overlap terms, and then estimate them using a quantum computer. In the second approach, we utilize the fact that the controller is related to the gradient of the Lyapunov function as mentioned in \cite{magann2022lyapunov} and use gradient evaluation techniques to evaluate the controller, such as the finite difference approximation method \cite{baydin2018automatic} and the parameter-shift-rule (PSR) \cite{wierichs2022general}. We demonstrate the algorithm through an illustrative example and through an application in quantum chemistry. We assess its performance, convergence properties, and robustness against sampling noise through extensive numerical simulations. Additionally, we demonstrate proof of principle execution on IBM's superconducting quantum computer.

The remainder of this paper is structured as follows. Section~2 reviews QLC and FALQON. Section~3 introduces the proposed FQAE and gives a detailed analysis of how to compute the controller using both classical and quantum computers in a hybrid manner. Section~4 investigates the efficiency of FQAE through an illustrative example and an application in quantum chemistry of computing the spectrum of the hydrogen molecule Hamiltonian. Finally, the outlook and conclusion are given in Section~5.

\section{Preliminaries}
\label{sec:paperB:Preliminaries}

This section presents a summary of FALQON for preparing ground states and tackling QUBO problems \cite{magann2022lyapunov,magann2022feedback}. We first examine QLC and then elucidate its relationship with FALQON.

\subsection{Quantum Lyapunov Control}
$\quad \:$ Consider the Hilbert space $\mathcal{H}=\mathbb{C}^N$ along with its corresponding orthonormal basis $\left.\mathcal{A}=\{|q_k\rangle\right\}_{ k \in\left\{0, \ldots, N-1\}\right.}$. Subsequently, all operators will be represented in the basis $\mathcal{A}$.

Consider a quantum system whose dynamics are governed by the controlled time-dependent Schrödinger equation 
 \begin{equation}
    i|\dot{\psi}(t)\rangle=H(t)|\psi(t)\rangle, \quad H(t)=H_0+H_c(t)
    \label{PB:Model}
\end{equation}
where the Hamiltonian $H(t)$ is rescaled by $\hbar$, $H_c(t)=\sum_{l=1}^r H_l u^{(l)}(t)$, $\{u^{(l)}(t)\}_{l=1}^r$ represents the control inputs, and $H_0$ and $H_c$ are the drift and control Hamiltonian, respectively. The Hamiltonians $H_0$ and $H_l$ are assumed to be non-commuting and time-independent: $[H_0, H_l]\neq0$, $l=1,\dots,r$. Assume that for the simplicity of the analysis, we only have one control input in the control Hamiltonian, i.e.,
\begin{equation}
    H_c(t)=u(t)H_1.
\end{equation}
Hence, the model becomes the following:
\begin{equation}
    i|\dot{\psi}(t)\rangle=(H_0+u(t)H_1)|\psi(t)\rangle.
    \label{PB:Model1}
\end{equation}
FALQON strives to solve the following control problem. 
\\
\\
\textbf{Problem 1:} \textit{Given a Hamiltonian $H_0$, the main objective is to find a control law in a feedback form, $u(\ket{\psi(t)})$, that guarantees the convergence of the quantum system \eqref{PB:Model1}, from any initial state to the ground state of the Hamiltonian $H_0$, i.e., the state $\ket{\psi_f}=\text{argmin}_{\ket{\psi}\in \mathcal{H}} \bra{\psi}H_0\ket{\psi}$.} 
\\ 
\\
We choose the matrix $H_0$ to be diagonal in the $\mathcal{A}$ basis with eigenvalues $ q_0< \dots<  q_{N-1}$ and corresponding eigenvectors $\ket{ q_0}, \dots, \ket{ q_{N-1}}$, and introduce our assumptions as follows: 
\\
\\
\textbf{\textit{Assumption 1.} } The drift Hamiltonian $H_0$ has distinct eigenvalue gaps, i.e., $ q_i- q_j \neq  q_k- q_l$ for all $(i,j) \neq (k,l).$
\\
\textbf{\textit{Assumption 2.} } The control Hamiltonian $H_1$ is fully connected, which means that all off-diagonal elements of $H_1$ are non-zero, i.e., $\bra{ q_i} H_1 \ket{ q_j} \neq 0$ for all $i \neq j$.
\\
\\
Consider a Lyapunov function of the form
\begin{equation}
    V(\ket{\psi(t)}) = \bra{\psi(t)}H_0\ket{\psi(t)}.
    \label{V1}
\end{equation}
The derivative of $V(\ket{\psi(t)})$ along the trajectories of system \eqref{PB:Model} is given by
\begin{align}
\dot{V}(\ket{\psi(t)}) & =  \bra{\psi(t)}  \mathrm{i}[H_1,H_0] \ket{\psi(t)} u(t) .
\label{vdot1}
\end{align}
Designing $u(t)$ as 
\begin{equation}
    u(t)= - K   g(\bra{\psi(t)}  \mathrm{i}[H_1,H_0] \ket{\psi(t)}),
    \label{PB:controller1}
\end{equation}
where $K>0$ and $g$ represents a continuous function that meets the conditions $g(0) = 0$ and $xg(x) > 0$ for all $x \neq 0$, ensures the condition $\dot{V}\leq0$. It is known in the literature (see, e.g., \cite{cong2013survey,grivopoulos2003lyapunov}) that applying the controller given by \eqref{PB:controller1} to the system \eqref{PB:Model1} guarantees asymptotic convergence of almost all initial states to the ground state of the Hamiltonian $H_0$ given Assumptions~1-2 are satisfied. Thus, designing the controller as \eqref{PB:controller1} solves Problem~1.

This Lyapunov control setup is proposed to operate on a general problem Hamiltonian $H_0$. Although, the assumptions introduced in Section~II (namely, non-degenerate spectral gaps for $H_0$ and full connectivity of $H_1$) are sufficient to guarantee convergence under the Lyapunov control setup, they are not strictly necessary in practice (see Appendix~A of \cite{magann2022lyapunov} for further discussion). In fact, in many practical problems, the condition on the mixer Hamiltonian (Assumption~2) cannot be satisfied. Our numerical results, together with earlier studies \cite{magann2022lyapunov,magann2022feedback,larsen2024feedback,malla2024feedback}, indicate that the algorithm retains good performance even when these conditions are not strictly satisfied. Several techniques can further enhance convergence without strict adherence to the assumptions, including the heuristic approaches outlined in the appendix of \cite{magann2022feedback}, the adaptive randomized strategy of \cite{magann2023randomized}, and the recent refinements reported in \cite{malla2024feedback,brady2025feedback,tang2025nonvariational}.

\subsection{Feedback-Based Quantum Optimization Algorithm} \label{SFQA}
We now demonstrate the connection between FALQON and continuous QLC discussed in the preceding subsection, in line with the content presented in \cite{magann2022lyapunov,magann2022feedback}. 
We start by considering the quantum evolution propagator $U(t)$, which is defined as the solution to Eq. \eqref{PB:Model} as follows:
\begin{equation}
    U(t)=\tau e^{-i \int_{0}^{t} H(s) d s},
    \label{propagator}
\end{equation}
where $\tau$ is the time-ordering operator. By decomposing it into $p$ piece-wise constant time intervals of length $\Delta t$, such that the total evolution time $T=p\Delta t$ we obtain
\begin{equation}
   U(T, 0) \approx \prod_{k=1}^{p} e^{-i H(k \Delta t) \Delta t},
   \label{U(T,0)}
\end{equation}
where $\Delta t$ is chosen sufficiently small such that $H(t)$ remains approximately constant within each interval $\Delta t$. By Trotterization, we can simplify this further to:

\begin{equation}
U(T, 0)  \approx \prod_{k=1}^{p} \bigl(e^{-i u(k \Delta t)  H_{1}\Delta t}  e^{-iH_{0}\Delta t} \bigr).
\end{equation}
Therefore, we obtain a digitized representation of the evolution in the following form.
\begin{align}
\ket{\psi_p}   &=\prod_{k=1}^{p} \bigl( e^{-i u(k \Delta t)  H_{1}\Delta t}e^{-iH_{0}\Delta t} \bigr) \ket{\psi_0} \nonumber \\
&=  \prod_{k=1}^{p} \bigl(U_1(u_k)U_0 \bigr) \ket{\psi_0},
\label{PB:evolution}
\end{align}
where we use the notation $u_k=u(k \Delta t)$, $\ket{\psi_k}=\ket{\psi(k \Delta t)}$, $U_0=e^{-iH_{0}\Delta t}$, $U_1(u_k)=e^{-i u(k \Delta t)  H_{1}\Delta t}$. Note that in each discrete time step, the controller's values correspond to the parameters of the circuit, and we will use these terms interchangeably.

\begin{figure}[H]
      \centering
       \captionsetup{justification=centering}
      \includegraphics[width=0.8 \linewidth]{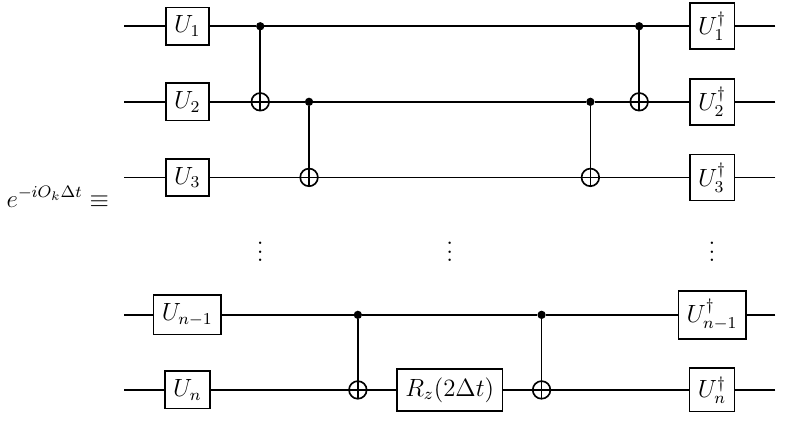}
      \caption{Quantum circuit implementation of the Pauli gadget for implementing the unitary evolution $e^{-iO_{k} \Delta t}$, where $O_{k}=O_{k,1} \otimes O_{k,2} \otimes \cdots \otimes O_{k,n}$ is a Pauli string. 
      and $U_{l}=\left\{\begin{array}{ll}
      R_y(-\pi/2) , & \text { if } O_{k,l}=X, \\
      R_x(\pi/2), & \text { if } O_{k,l}=Y, \\
      I, & \text { if } O_{k,l}=Z.
      \end{array} \right.$}
      \label{PB:circuit}
   \end{figure}

The unitary $U_0$ can be efficiently implemented as a quantum circuit given the drift Hamiltonian $H_0$ is specified as a sum of Pauli strings $H_0=\sum_{k=1}^{N_0} c_{k} O_{k}$, where $N_0$ is given as a polynomial function of the number of qubits $n$, $\{c_{k}\}_{k=1}^{N_0}$ are real scalar coefficients,  and a Pauli string is a Hermitian operator in the form $O_{k}=O_{k,1} \otimes O_{k,2} \otimes \cdots \otimes O_{k,n}$ with $O_{k,l} \in \{ I,X,Y,Z \}$. Furthermore, to be able to implement unitary $U_1(u_k)$ as a quantum circuit efficiently, the Hamiltonian $H_1$ is designed as a linear combination of Pauli strings as $ H_1 = \sum_{k=1}^{N_1} \Bar{c}_k \Bar{O}_k $, where $N_1$ is a polynomial function of the number of qubits. For such a control Hamiltonian design, the Hamiltonians $H_0$ and $H_1$ are decomposed in terms of a polynomial number of Pauli strings. Therefore, the quantum circuit that implements the evolution defined in \eqref{PB:evolution} can be efficiently implemented on a quantum computer. For a detailed analysis of the implementation of the quantum circuit of unitaries $U_0$ and $U_1(u_k)$, we refer the reader to \cite{magann2021digital}. Figure~\ref{PB:circuit} shows the circuit implementation of a unitary in the general form of $e^{-iO_{k} \Delta t}$. This circuit can simply be generalized to the case of a linear combination of Pauli strings (see \cite{magann2021digital}).

Including multiple controllers often improves convergence, as noted in Section 3.C of Magann et al.\cite{magann2022lyapunov}. A physically motivated way to construct the control Hamiltonian $H_c$ has been proposed by Malla et al. \cite{malla2024feedback}, where the design of the control Hamiltonian is inspired by the counterdiabatic driving protocol rooted in quantum adiabaticity. While we heuristically design the control Hamiltonians in our work, a systematic design of the control Hamiltonian given an arbitrary Hamiltonian $H_0$ remains an open challenge.

We have so far approximated the continuous-time evolution by a sequential product of discrete time steps, and the resulting quantum circuit that implements $U(T,0)$ simulates the propagator represented by Eq. \eqref{propagator}, where the condition $\dot{V}\leq0$ can be guaranteed to hold by choosing $\Delta t$ to be sufficiently small. For a detailed analysis of how to choose $\Delta t$, see Appendix A in \cite{magann2022lyapunov}. For the control law, a discrete version of the controller given by \eqref{PB:controller1} is adopted as follows:
\begin{equation}
    u_{k+1}= -K    \bra{\psi_k} \mathrm{i} [H_1,H_0]  \ket{\psi_k},
    \label{PB:udis1}
\end{equation}
where the function $g(\cdot)$ is chosen as the identity function.

To evaluate the controller, we need to estimate the term $\bra{\psi_k}  \mathrm{i}[H_1,H_0]  \ket{\psi_k}$. Expanding this term by means of Pauli strings, we get
\begin{equation}
    \label{PB:expectation}
     \bra{\psi_k}  \mathrm{i}[H_1,H_0]  \ket{\psi_k} = \sum_{j=1}^{N_2} \hat{c}_q \bra{\psi_k} \hat{O}_{j}
     \ket{\psi_k},
\end{equation}
where we employed the product rule of Pauli strings, which asserts that $O^jO^k=\delta_{j,k}I + \mathrm{i} \epsilon_{jkl} O^l$, where $O^1=X$, $O^2=Y$ and $O^3=Z$. Here, $\delta_{jk}$ represents the Kronecker delta, and $\epsilon_{jkl}$ denotes the Levi Civita symbol. Therefore, the value of $\bra{\psi_k}  \mathrm{i}[H_1,H_0]  \ket{\psi_k}$ can be determined by estimating the expectations of $N_2$ Pauli strings, where the specific value of $N_2$ is determined by the characteristics of the Hamiltonians $H_0$ and $H_1$. Since $N_0$ and $N_1$ are polynomial functions of the number of qubits, $N_2$ will also be a polynomial function of the number of qubits. Note that the scalling of $N_2$ could be large. For example, as pointed out by an anonymous reviewer, for the electronic structure problem with a control Hamiltonian that scales as $n^d$, in the worst case, the number of Pauli terms in the commutator $i[H_1, H_0]$ could scale as $n^{(4+d)}$. Despite this large scaling, this quantity remains polynomial in the number of qubits. In such cases, although the total number of terms may be large, more efficient quantum measurement strategies can be employed to address the associated measurement overhead. Techniques such as adaptive informationally complete measurements \cite{garcia2021learning}, shadow measurements \cite{bertuzzi2025shadow}, and grouping of compatible commuting observables \cite{yen2020measuring} can be employed to significantly reduce the measurement cost associated with estimating expectations of large Pauli sums. In particular, the study by Bertuzzi et al. \cite{bertuzzi2025shadow} demonstrates how shadow measurements can be adapted to the feedback-based setup to significantly reduce the number of measurements to estimate the observables required for calculating the feedback-based controller.

\begin{figure}[H]
    \centering
    \includegraphics[ width=0.95 \linewidth]{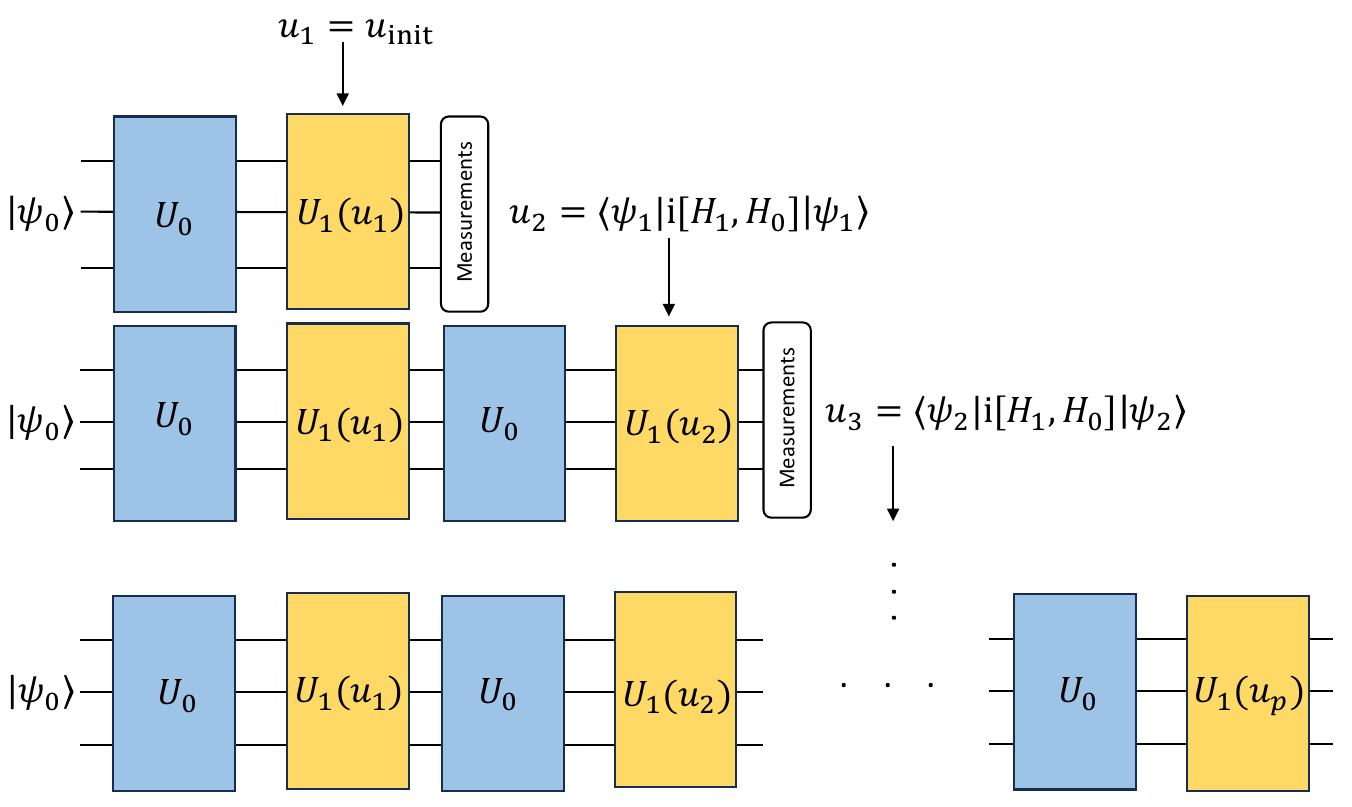}
    \caption{This figure is adapted from \cite{magann2022feedback}. The figure demonstrates the procedural steps involved in FALQON. In the first step, the algorithm is fed with an initial guess of the value of the controller, the state $\ket{\psi_1}$ is prepared in the quantum computer and the controller for the next layer $u_2$ is calculated. Over the following $k$ iterations, the quantum circuit expands by adding a layer composed of $U_1(u_k)U_0$, with $u_k$ calculated from the preceding layer using Equation \eqref{PB:udis1}. The process concludes upon reaching the maximum layer count $p-1$.}  
    \label{PB:FQAalg}
\end{figure}

The procedure for implementing FALQON is detailed in Algorithm~1 and depicted in Figure \ref{PB:FQAalg}. The time step is initially set to $\Delta t$, and the procedure is seeded with an initial value of $u_1 = u_{\text{init}}$, which is set to $ u_{\text{init}}=0$. Following this, a group of qubits is initialized to an easy-to-prepare initial state $\ket{\psi_0}$, and a single circuit layer is applied to prepare the state $\ket{\psi_1}$. Subsequently, the controller $u_2$ for the next layer of the quantum circuit is estimated using the quantum computer. Next, an additional layer is added to the circuit, and this process is repeated iteratively for a total of p layers. The resulting iteratively designed quantum circuit to approximate the ground state $\prod_{k=1}^{p} \bigl(U_1(u_k)U_0 \bigr)$, together with its parameters $\{u_k\}_{k=1}^p$, forms the output of the algorithm.
\algnewcommand\algorithmicInput{\textbf{Input:}}
\algnewcommand\algorithmicOutput{\textbf{Output:}}
\algnewcommand\Input{\item[\algorithmicInput]}
\algnewcommand\Output{\item[\algorithmicOutput]}
   
\setcounter{algorithm}{0}   

\begin{algorithm}[H]
\caption{FALQON \cite{magann2022lyapunov}}\label{PB:FALQON}
\begin{algorithmic}[1]
\Input{$H_0$, $H_1$, $\Delta t$, $p$, $\ket{\psi_0}$}
\Output{Quantum Circuit parameters $\{u_k\}_{k=1}^p$}
\State Set $u_1=0$
\State{\textbf{for} $k = 1$ to $p-1$ \textbf{do}}
\State Prepare the initial state $\ket{\psi_0}$
\State Prepare the state $\ket{\psi_k}=\prod_{l=1}^{k} \bigl(U_1(u_l)U_0 \bigr)\ket{\psi_0}$
\State {Estimate the term $\bra{\psi_k}  \mathrm{i}[H_1,H_0]  \ket{\psi_k}$ by estimating each of the expectations in \eqref{PB:expectation} on the quantum computer}
\State {Calculate $u_{k+1}$ using \eqref{PB:udis1} }
\State{\textbf{end for}}
\end{algorithmic}
\end{algorithm}

\section{Feedback-Based Quantum Optimization Algorithm for Excited States Calculation}
\label{sec:paperB:Feedback-Based Quantum Optimization Algorithm for Excited States Calculation}

In this section, we introduce the problem of calculating excited states of a given Hamiltonian based on QLC and propose an approach to solving this problem by defining a new operator that encodes the m'th excited state of the original Hamiltonian as its ground state. We then design the control law such that the trajectories converge to this operator's ground state. We demonstrate that employing this method enables an efficient implementation of the quantum circuit and facilitates efficient controller computation.

We start by considering the same model as \eqref{PB:Model1}. In this model, we assumed that we have one control input. This assumption can be extended to the case of multiple control inputs, as given in the appendix. We strive to solve the following control problem. 
\\ 
\\
\textbf{Problem 2:} \textit{Given a Hamiltonian $H_0$ 
 with its $m-1$ lowest energy eigenstates
$\{\ket{q_0}, ..., \ket{q_{m-1}}\}$, find a control law in a feedback form, $u(\bra{\psi(t)} A \ket{\psi(t)})$ where $A$ is an observable, that assures the convergence of the quantum system \eqref{PB:Model1}, from any initial state to the m'th excited state of the Hamiltonian $H_0$.} 
\\ 
\\
\noindent
Note that in Problem~2 we assume the lower energy eigenstates are known. 

We assume that Assumption~1-2 hold. Note that the controller's design is constrained to the form of an expected value of an observable $u(\bra{\psi(t)} A \ket{\psi(t)})$ to facilitate efficient evaluation using quantum computers. \\
We consider a Lyapunov function of the form
\begin{equation}
    V(\ket{\psi}) = \bra{\psi(t)}H_0\ket{\psi(t)} + \sum_{j=0}^{m-1} \alpha_j \abs{\braket{\psi(t)}{q_j}}^2,
    \label{V}
\end{equation}
where $\{\ket{q_0}, ..., \ket{q_{m-1}}\}$ are the lower energy eigenstates of the Hamiltonian $H_0$, and $\{ \alpha_0, ..., \alpha_{m-1} \}$ are constants that should be chosen to be sufficiently large. More details on choosing these values will be provided subsequently. \\
Equation \eqref{V} can be expressed as
\begin{align}
    V(\ket{\psi(t)}) 
    &= \bra{\psi(t)}H_0\ket{\psi(t)} \nonumber +  \bra{\psi(t)} \sum_{j=0}^{m-1} \alpha_j \ketbra{q_j} \ket{\psi(t)}, \nonumber \\
    & = \bra{\psi(t)} \bigl(H_0 + \sum_{j=0}^{m-1} \alpha_j \ketbra{q_j} \bigr)\ket{\psi(t)}, \nonumber \\
    & = \bra{\psi(t)} P \ket{\psi(t)}.
    \label{LyapunovFunc}
\end{align}
Here, we introduced a new Hermitian operator $P$ defined as
\begin{equation}
    P := H_0+\sum_{j=0}^{m-1} \alpha_j M_j,
    \label{P}
\end{equation}
where $M_j=\ketbra{q_j}$ is the projector on the j'th excited state. Note that the operator $P$ commutes with $H_0$, i.e., $[P,H_0]=0$, hence, it is diagonal in the basis $\mathcal{A}$, with eigenvalues $p_0,p_1,...,p_{N-1}$. In addition, given that the parameters $\alpha_j$ are chosen sufficiently large ($\alpha_j > q_m - q_j$, $j=0,1,...,m-1$), the ground state of the operator $P$, i.e., $\ket{\psi_g}=\text{argmin}_{\ket{\psi}\in \mathcal{H}} \bra{\psi}P\ket{\psi}$ is the same as the m'th excited state of the Hamiltonian $H_0$. Therefore, we propose a new control problem equivalent to Problem~2 as follows. 
\\
\\
\textbf{Problem 3:} \textit{Given the Hamiltonian $H_0$ with its $m-1$ lowest energy eigenstates $\{\ket{q_0}, ..., \ket{q_{m-1}}\}$, find a control law in a feedback form $u(\bra{\psi(t)} A \ket{\psi(t)})$ where $A$ is an observable, that guarantees the convergence of the quantum system \eqref{PB:Model1}, from any initial state to the ground state of the operator $H_0 + \sum_{k=0}^{m-1} \alpha_j \ketbra{q_k}$.}
\\
\\
\noindent
The derivative of $V(\ket{\psi(t)})$ along the trajectories of system \eqref{PB:Model} is given by
\begin{align}
\dot{V}(\ket{\psi(t)}) & =  \bra{\psi(t)} \mathrm{i} [H_1,P] \ket{\psi(t)} u(t).
\label{vdot}
\end{align}
We design $u(t)$ such that $\dot{V}\leq0$
\begin{equation}
    u(t)= - K   g(\bra{\psi(t)}  \mathrm{i}[H_1,P] \ket{\psi(t)}),
    \label{PB:controller}
\end{equation}
where $K>0$ and $g$ represents a continuous function that meets the conditions $g(0) = 0$ and $xg(x) > 0$ for all $x \neq 0$.

Applying the controller given by \eqref{PB:controller} to the system \eqref{PB:Model1} ensures asymptotic convergence of almost all initial states to the ground state of the operator $P$, given in addition to Assumptions~1-2, the following assumption is also satisfied \cite{cong2013survey,grivopoulos2003lyapunov}:
\\
\\
\textbf{\textit{Assumption 3.}} The operator $P$ has a non-degenerate spectrum, i.e., $p_i \neq p_j$ for all $i \neq j$.
\\
\\
The choice of the shifting parameters $\{\alpha_j\}_{j=0}^{m-1}$ in the design of the operator $P$ is crucial to guarantee that Problem 3 is equivalent to Problem 2. In the following, we give a detailed analysis of the choice of the parameters $\{\alpha_j\}_{j=0}^{m-1}$. 

For the operator $P$ defined as in \eqref{P}, and for a general state expressed in the basis $\mathcal{A}$ as $\ket{\psi}=\sum_{k=0}^{N-1} \beta_k\ket{q_{k}}$, 
we have
\begin{equation}
\label{Pe}
    \expval{P}{\psi} = \sum_{k=0}^{m-1}\left|\beta_{k}\right|^2\left(q_k+\alpha_k\right)+\sum_{k=m}^{N-1}\left|\beta_k\right|^2 q_k,
\end{equation}
where $q_0, q_1, ..., q_{m-1} $ are the eigenvalues of the lower energy eigenstates. To guarantee that the ground state of the operator $P$ is the targeted m'th excited state of the Hamiltonian $H_0$, we need to choose $\alpha_k > q_m-q_k$ for all $k \in \{0, 1, \ldots, m-1\}$. Ideal performance is achieved by choosing these parameters large enough to make Problem~2 and Problem~3 equivalent and avoid excessively high values of $\alpha_{ k}$ as they may adversely affect the algorithm's efficiency.  In our analysis, we assume that the lower energy eigenstates are known, and accordingly, we can find their corresponding eigenvalues. However, the energy of the targeted eigenstate $q_m$ is unknown. Therefore, we cannot choose $\alpha_{ k}$ directly. In the following, we adopt three alternative approaches for choosing the shifting parameters $\alpha_{ k}$ similar to those proposed for designing the cost function for VQD algorithm \cite{higgott2019variational}.

The first approach is to find an upper bound on the energy gap $G$ defined as $G := q_{\max}-q_{\min} \geq q_m-q_{ k}$, where $q_{\max}$ and $q_{\min}$ are the maximum and minimum eigenvalues of $H_0$, respectively. For a drift Hamiltonian $H_0$ specified as a sum of Pauli strings $H_0=\sum_{{ k}=1}^{N_0} c_{ k}O_{ k}$, an upper bound can be found as 
\begin{equation}
    G \leq 2||H|| \leq 2 \sum|c_q|.
    \label{G}
\end{equation}
This upper limit provides insights into selecting the appropriate value for $\alpha_{ k}$.

In the second approach, we possess an accurate estimate of the gap $G$ by evaluating $q_{\min}$ and $q_{\max}$ using a quantum algorithm, where $-H_0$ is used to evaluate $q_{\max}$.

The third approach involves an iterative adjustment of the parameters $\alpha_j$. We initialize $\alpha_{ k}$ with a small guess, and if convergence to the lower energy eigenstate is observed, indicating that $\alpha_{ k} < q_m - q_{ k}$, we double its value and repeat the process. This iterative procedure guarantees $\alpha_{ k} > q_m - q_{ k}$ after $O(\log_2(q_m - q_{k}))$ repetitions of the algorithm. Although this approach requires multiple evaluations, it can offer improved performance by choosing smaller parameter values.

Following the same Trotterization procedure presented in Subsection~\ref{SFQA}, we get the same quantum circuit as in \eqref{PB:evolution}. For the controller, we adopt the following feedback law:
\begin{equation}
    u_{k+1}= -K    \bra{\psi_k} \mathrm{i} [H_1,P]  \ket{\psi_k},
    \label{PB:udis}
\end{equation}
which represents a discrete version of the controller given by Eq. \eqref{PB:controller}, and the function $g(\cdot)$ is chosen as the identity function.

Since FQAE has the same form of the quantum circuit as FALQON, the construction of the quantum circuit of FQAE follows the same procedures as for FALQON presented in Subsection \ref{SFQA}. However, the controller evaluation is different from that of FALQON. Since the operator $P$ cannot be efficiently expressed in terms of Pauli strings, we will not expand the term $\bra{\psi_k} \mathrm{i} [H_1,P]  \ket{\psi_k}$ in a Pauli strings' basis and therefore we cannot directly evaluate the controller given by \eqref{PB:udis} using the same procedure as in FALQON. In the following subsection, we give different approaches on how to evaluate the controller in a hybrid way using both quantum and classical computers. The details of the algorithmic steps to implement FQAE are given in Algorithm 2.

    \begin{algorithm} [H]
    \caption{FQAE}\label{FQAE}
    \begin{algorithmic}[1]
    \Input{$H_0$, $H_1$, $\Delta t$, $p$, $\ket{\psi_0}$, $\{\alpha_j\}_{j=1}^{m-1}$}
    \Output{circuit parameters $\{u_k\}_{k=1}^p$}
    \State{Design the operator $P$ using Eq. \eqref{P}}
    \State{Set $u_1=u_\text{init}$}
    \State{\textbf{for} $k = 1$ to $p-1$ \textbf{do}}
    \State Prepare the initial state $\ket{\psi_0}$
    \State Prepare the state $\ket{\psi_k}=\prod_{l=1}^{k} \bigl(U_1(u_l)U_0 \bigr)\ket{\psi_0}$
    \State {Calculate the controller for the next layer $u_{k+1}$ using one of the approaches detailed in Subsection~\ref{cc}. }
    \State{\textbf{end for}}
    \end{algorithmic}
    \end{algorithm}
\noindent
\textbf{Remark 1.}
Our analysis assumes that the lower energy eigenstates, their eigenvalues, and the circuits that prepare them are known. If the lower energy eigenstates are not known, we can use FQAE in an iterative manner by calculating the lowest energy eigenstate $\ket{q_{ 0}}$. Subsequently, utilizing $\ket{q_0}$, we can calculate $\ket{q_1}$. Repeating this process for all lower energy eigenstates, we can evaluate the $ m$'th excited state. 
\\ 
\\
\textbf{Remark 2.}
In addition to being an easy-to-prepare state, various choices for the initial state could be employed to enhance the algorithm's performance. For instance, in \cite{larsen2024feedback}, the initial state $\ket{\psi_0}$ was selected as the ground state of the control Hamiltonian $H_1$, similar to quantum annealing. Another choice involves utilizing a warm-starting technique such as \cite{egger2021warm}, where the initial state corresponds to the solution of a relaxed combinatorial optimization problem. Alternatively, one could adopt a VQA with a short-depth circuit, using its output as the initial state. We defer a comprehensive analysis of different initial state choices to future work. 
\\
\\
\textbf{Remark 3. }When seeking to compute the ground state of the Hamiltonian $H_0$, since there are no eigenstates with lower energy, according to equation \eqref{P}, we obtain $P=H_0$. This design of the operator $P$ leads to the recovery of FQA. Consequently, FQA can be regarded as a special case of FQAE, wherein the operator $P$ is defined as $H_0$, and the targeted eigenstate is the ground state of the Hamiltonian $H_0$. 
\\
\\
\textbf{Remark 4.} FALQON can be directly applied to our problem once the operator $P$ in \eqref{P} is used as a new problem Hamiltonian that encodes the m'th excited state of the Hamiltonian $H_0$ as its ground state. However, such a direct application of FALQON differs from our proposed approach FQAE. As shown in Section~2, in FALQON (Algorithm~1), this new Hamiltonian $P$ will be used both in the controller calculation and in the quantum circuit implementation. Specifically, we will need to implement the unitary $e^{-i  P\Delta t}$, which will result in a deep quantum circuit, as this will require the implementation of the projectors as quantum circuits. On the other hand, in FQAE, we design the operator $P$ and use it only in the controller estimation. According to the Lyapunov control setup, convergence to the ground state of the operator $P$ (the $m$'th excited state of $H_0$) is still achieved when the operator $P$ is only used in the Lyapunov function while $H_0$ remains as the drift part to be realized by a quantum circuit. This result was first rigorously established by Grivopoulos and Bamieh \cite{grivopoulos2003lyapunov} and has since been further developed in several subsequent works (see, for example, \cite{cong2013survey,clausen2023measurement}). Under this setup, there is no need to implement the unitary $e^{-i P \Delta t}$, which would otherwise result in significantly deeper quantum circuits. 

\subsection{Controller Computation} \label{cc}
This subsection provides a detailed analysis of the controller's evaluation using a hybrid approach with quantum and classical computers. We propose two alternative methods for evaluating the controller.

\subsubsection{Expectation and Overlap Estimation-Based Approach for Evaluating the Controller}
\begin{figure}[H]
    \centering
    \includegraphics[ width=0.9 \linewidth]{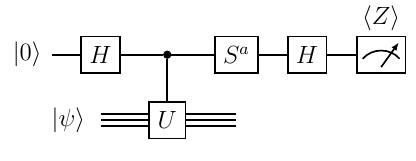}
    \caption{Basic Hadamard test, where $a \in \{0, 1\}$, $H$ is the Hadamard gate, $S=e^{-i\pi Z/4}$ is the phase gate, and $U$ is the unitary we are interested in its expected value with respect to the state $\ket{\psi}$. For $a = 0$, $\expval{Z}$ corresponds to the real part of the inner product $\text{Re}\bra{\psi} U \ket{\psi}$, and for $a = 1$, $\expval{Z}$ corresponds to the imaginary part of the same inner product $\text{Im}\bra{\psi} U \ket{\psi}$.}
    \label{HT}
\end{figure}

We introduce our first approach for evaluating the controller. In this approach, we reduce the evaluation of the controller given by \eqref{PB:udis} into several terms, including expected values of the form $\bra{\psi} B \ket{\psi}$ and overlap terms of the form $\bra{\psi} C \ket{\phi}$ where $B$ and $C$ are Pauli strings, and $\ket{\psi}$ and $\ket{\phi}$ are arbitrary quantum states. We estimate these terms on a quantum computer and then use these estimates to compute the controller by post-processing on a classical computer. To evaluate the overlap terms, we use the Hadamard test \cite{aharonov2006polynomial}. The Hadamard test serves as a subroutine in quantum computing that enables estimation of the real and imaginary parts of the inner product $\bra{\psi} U \ket{\psi}$ for an arbitrary quantum state $\ket{\psi}$ and unitary $U$. The quantum circuit for the Hadamard test is depicted in Figure~\ref{HT}.

\newpage
For the first method, using the controller defined in \eqref{PB:udis} and the operator $P$ defined in \eqref{P} and using the bilinearity property of the commutator, we obtain:
\begin{align}
    & u_{k+1} \nonumber \\
    & = -K   \bra{\psi_k}  \mathrm{i} [H_1,P]  \ket{\psi_k}, \nonumber \\
            & = -K    \bra{\psi_k} \mathrm{i}  [H_1,H_0+\sum_{j=0}^{m-1} \alpha_j M_j]  \ket{\psi_k}, \nonumber \\
            & = -K    \Bigl( \bra{\psi_k}  \mathrm{i}[H_1,H_0]  \ket{\psi_k} +  \bra{\psi_k}  \mathrm{i} [H_1,\sum_{j=0}^{m-1} \alpha_j M_j]  \ket{\psi_k} \Bigr), \nonumber \\
         &= -K    \Bigl( \bra{\psi_k}  \mathrm{i}[H_1,H_0]  \ket{\psi_k} + \bra{\psi_k}  \mathrm{i} (H_1 \sum_{j=0}^{m-1} \alpha_j M_j)\ket{\psi_k}  - \bra{\psi_k}  \mathrm{i} ((\sum_{j=0}^{m-1} \alpha_j M_j) H_1)  \ket{\psi_k} \Bigr), \nonumber \\
         &= -K    \Bigl( \bra{\psi_k}  \mathrm{i}[H_1,H_0]  \ket{\psi_k} +\bra{\psi_k}  \mathrm{i} (H_1 \sum_{j=0}^{m-1} \alpha_j M_j)\ket{\psi_k} \nonumber +(\bra{\psi_k}  \mathrm{i} (H_1 \sum_{j=0}^{m-1} \alpha_j M_j)\ket{\psi_k})^* \Bigr), \nonumber\\
         &= -K    \Bigl( \bra{\psi_k}  \mathrm{i}[H_1,H_0]  \ket{\psi_k} +2 \cdot \text{Re}\{\bra{\psi_k}  \mathrm{i} (H_1 \sum_{j=0}^{m-1} \alpha_j M_j)\ket{\psi_k}\}\Bigr),
         \nonumber \\
         &= -K    \Bigl( \bra{\psi_k}  \mathrm{i}[H_1,H_0]  \ket{\psi_k} +2\cdot\text{Re}\{ \mathrm{i}\cdot\sum_{j=0}^{m-1} \alpha_j \bra{\psi_k} H_1 \ket{q_j} \braket{q_j}{\psi_k}\}\Bigr).
    \label{PB:uest}
\end{align}
As outlined in the following steps, we estimate the terms $\bra{\psi_k}  \mathrm{i}[H_1,H_0]  \ket{\psi_k}$ , $\bra{\psi_k} H_1 \ket{q_j}$ and $\braket{q_j}{\psi_k}$ using hybrid quantum and classical processors.
\begin{enumerate}
    \item Estimating $\bra{\psi_k}  \mathrm{i}[H_1,H_0]  \ket{\psi_k}$: \\
     Expanding this term in terms of Pauli strings, we get
\begin{equation}
    \label{vk1}
     \bra{\psi_k}  \mathrm{i}[H_1,H_0]  \ket{\psi_k} = \sum_{q=1}^{N_2} \hat{c}_q \bra{\psi_k} \hat{O}_q \ket{\psi_k}.
\end{equation}
Therefore, by estimating expectations of $N_2$ Pauli strings, we can estimate the term $\bra{\psi_k}  \mathrm{i}[H_1,H_0]  \ket{\psi_k}$. The quantum circuit used to estimate this term is shown in Figure~\ref{sub1}. 
\item Estimating $\bra{\psi_k} H_1 \ket{q_j}$ for all $j \in \{0,1, ..., m-1\}$:

To estimate $ \bra{\psi_k} H_1 \ket{q_j}$, we decompose this term as
\begin{align}
\label{cB}
    \bra{\psi_k} H_1 \ket{q_j} &=  \bra{0} U_k^\dagger (\sum_{q=1}^{N_1} \Bar{c}_q \Bar{O}_q) U^{(j)} \ket{0} \nonumber \\
    &= \sum_{q=1}^{N_1} \Bar c_q \bra{0} U_k^\dagger \Bar{O}_q U^{(j)} \ket{0}, 
\end{align}
where $U_k$ is the circuit to prepare the state $\ket{\psi_k}=U_k \ket{0}$, $U^{(j)}$ is the circuit to prepare the state $\ket{q_j}=U^{(j)} \ket{0}$ (See Remark~1). Each of the terms in Eq. \eqref{cB} is then estimated on the quantum computer using the Hadamard test shown in Figure~\ref{sub2}. 

\item Estimating $\braket{q_j}{\psi_k}$ for all $j \in \{0,1, ..., m-1\}$:

The term $\braket{q_j}{\psi_k}$ can be estimated using the Hadamard test shown in Figure~\ref{sub3}, as follows:
\begin{align}
    \braket{q_j}{\psi_k} &= \bra{0} U^{(j) \dagger} U_k \ket{0}.
\end{align}

\begin{figure}[H]
    \centering
    
    \begin{subfigure}{\columnwidth}
        \centering
        \includegraphics[ width=0.55 \linewidth]{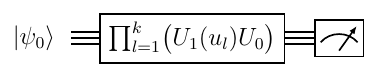}

        \caption{The quantum circuit used to estimate the term $\bra{\psi_k}  \mathrm{i}[H_1,H_0]  \ket{\psi_k}$.}
        \label{sub1}
    \end{subfigure}
    
    \begin{subfigure}{\columnwidth}
        \centering
        \includegraphics[ width=0.68 \linewidth]{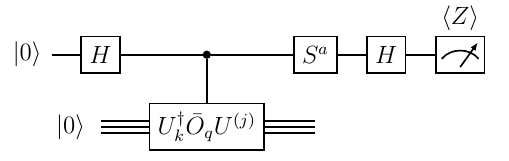}
 	 \caption{The Hadamard test circuit to estimate the term $\bra{0} U_k^\dagger \Bar{O}_q U^{(j)} \ket{0}$. }
        \label{sub2}
    \end{subfigure}

    \begin{subfigure}{\columnwidth}
        \centering
        \includegraphics[ width=0.68\linewidth]{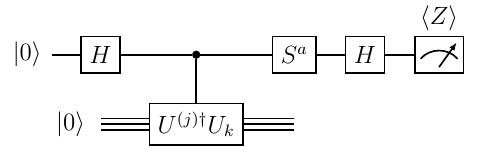}
        \caption{The Hadamard test circuit to estimate the term $\bra{0} U^{(j) \dagger} U_k \ket{0}$. }
        \label{sub3}
    \end{subfigure}

    \caption{The quantum circuits used to estimate the different terms of the controller $u_{k+1}$.}
    \label{A1}
\end{figure}

\item After estimating the term $\bra{\psi_k}  \mathrm{i}[H_1,H_0]  \ket{\psi_k}$, and the terms  $\bra{\psi_k} H_1 \ket{q_j}$ and $\braket{q_j}{\psi_k}$ for all $j \in \{0,1, ..., m-1\}$ using the quantum computer, these values are fed to a classical computer to compute the controller \eqref{PB:uest} as a classical post-processing step. A schematic of the first approach is given in Figure~\ref{FA}.

\begin{figure}[H]
    \centering
    \includegraphics[ width=1 \linewidth]{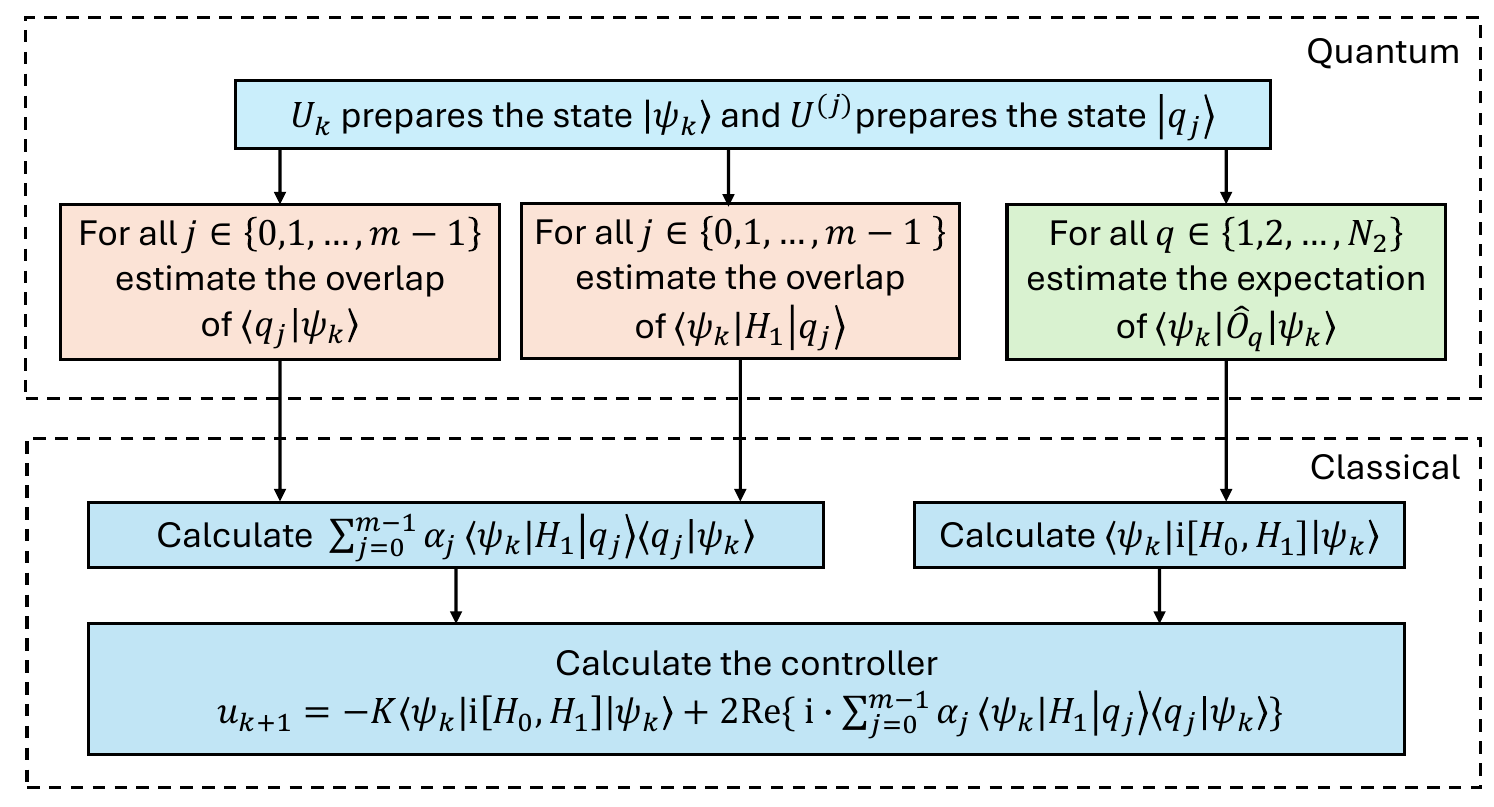}
    \caption{The expectation and overlap estimation-based approach for calculating the controller. }
    \label{FA}
\end{figure}

\end{enumerate}
In the subsequent subsection, we propose an alternative approach that circumvents the use of Hadamard tests, which can offer some advantages as will be discussed in the next subsection.

\subsubsection{Gradient-Based Approach for Evaluating the Controller}
As shown in the previous subsection, evaluating the controller involves several Hadamard tests, which require controlled versions of the unitaries that result in deep circuits, making the approach challenging for the current NISQ devices. In this subsection, we introduce an alternative method for evaluating the controller by establishing a connection between the controller and the gradient of the Lyapunov function. We illustrate the relationship between the controller and the Lyapunov function's gradient and propose two evaluation methods: a finite difference approximation and an analytical approach utilizing PSR.

At the k'th step of Algorithm \ref{FQAE}, the algorithm starts from the initial state $\ket{\psi_0}$ and $k$ layers are applied to get the state
\begin{align}
    \ket{\psi_k} &= \prod_{l=1}^{k} (U_1(u_l)U_0)\ket{\psi_0} \nonumber \\
    &= U_1(u_k)U_0 \prod_{l=1}^{k-1} (U_1(u_l)U_0)\ket{\psi_0}
    \label{l-1}.
\end{align}
From \eqref{LyapunovFunc} and using \eqref{l-1}, we let $\ket{\psi_0}$ and $u_k$ vary and fix $u_1,u_2,...,u_{k-1}$, hence we get
\begin{align}
   V_k(\ket{\psi_0},u_k)     &  \hat= \bra{\psi_k} P \ket{\psi_k}, \nonumber \\
                        &=        \bra{\psi_0} \Bigg( \Bigl(\prod_{l=1}^{k-1}U_0^\dagger U_1(u_l)^\dagger \Bigr) U_0^\dagger U_1(u_k)^\dagger  P  U_1(u_k)U_0 \Bigl(\prod_{l=1}^{k-1} U_1(u_l)U_0 \Bigr) \Bigg)\ket{\psi_0}, \nonumber \\
                        &= \bra{\psi_{k-1}} U_0^\dagger U_1(u_k)^\dagger P U_1(u_k)U_0 \ket{\psi_{k-1}}, \nonumber \\
                        &= \bra{\varphi} U_1(u_k)^\dagger P U_1(u_k) \ket{\varphi},
\end{align}
where $\ket{\varphi}=U_0 \ket{\psi_{k-1}}$. \\
For the controlled unitary $U_1(u_k) = e^{-i\Delta t{u_k} H_1}$, the gradient of this unitary with respect to the controller $u_k$ is given by:
\begin{align}
      \frac{\partial}{\partial u_k} U_1(u_k) &= -\mathrm{i} \Delta t H_1 U_1(u_k).
      \label{dUdu}
\end{align}
Hence, the gradient of $V$ with respect to $u_k$, and using \eqref{dUdu}, is given as:
\begin{align}
    \frac{\partial}{\partial u_k} V_k(u_k)     &=  \frac{d}{du_k} \langle \phi| U_1^\dagger(u_k) P U_1(u_k) |\phi \rangle, \nonumber \\
    &= \mathrm{i} \Delta t \langle \phi| \big( U_1^\dagger(u_k) H_1 P U_1(u_k) - U_1^\dagger(u_k) P H_1 U_1(u_k) \big) |\phi \rangle, \nonumber \\
    &= \mathrm{i} \Delta t \langle \phi| U_1^\dagger(u_k) (H_1 P - P H_1) U_1(u_k) | \phi \rangle, \nonumber \\
    &= \mathrm{i} \Delta t \langle \phi| U_1^\dagger(u_k) [H_1,P] U_1(u_k) | \phi \rangle, \nonumber \\
    &= \mathrm{i} \Delta t \bra{\psi_{k-1}} U_0^\dagger U_1^\dagger(u_k) [H_1,P] U_1(u_k) U_0  \ket{\psi_{k-1}}, \nonumber \\
    &=  \Delta t \bra{\psi_{k}}  \mathrm{i}[H_1,P]  \ket{\psi_{k}}.
    \label{dVdu}
\end{align}
From Equations \eqref{PB:udis} and \eqref{dVdu}, we get 
\begin{align}
    \label{PB:u_dv}
     u_{k+1} &= - \frac{K}{\Delta t} \frac{\partial}{\partial u_k} V_k(u_k), \nonumber \\
         & = - \frac{K}{\Delta t}( \frac{\partial}{\partial u_k} \bra{\psi_k}H_0\ket{\psi_k} + \frac{\partial}{\partial u_k} \sum_{j=0}^{m-1} \alpha_j \abs{\braket{\psi_k}{q_j}}^2 ).
\end{align}

\begin{figure}[H]
    \centering
    \includegraphics[ width=1 \linewidth]{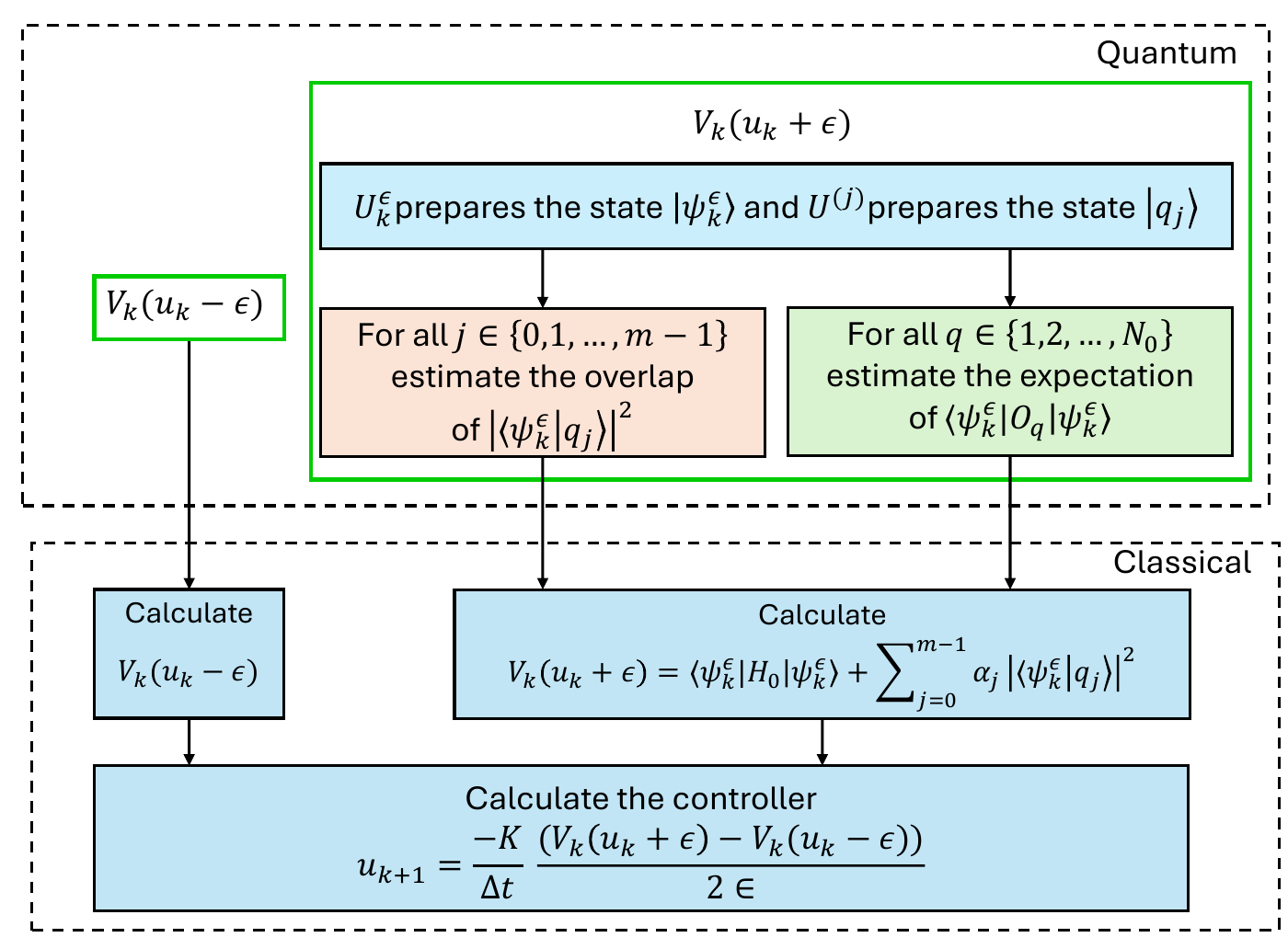}
    \caption{The gradient-based approach for calculating the controller using finite difference approximation to evaluate the gradient.}
    \label{FD}
\end{figure}

Therefore, to evaluate the controller for the next layer, we need to evaluate $\frac{\partial}{\partial u_k} V_k(u_k)$, which can be evaluated numerically on the quantum computer using the finite difference approximation of the gradient as follows \cite{baydin2018automatic}:
\begin{equation}
   \frac{\partial}{\partial u_k} V_k(u_k) \approx \frac{V_k\left(\boldsymbol{u_k}+\epsilon \right)-V_k\left(\boldsymbol{u_k}-\epsilon\right)}{2 \epsilon}.
   \label{PB:FDa}
\end{equation}
This means we need to evaluate the Lyapunov function twice to evaluate the controller. A schematic diagram of the gradient-based approach utilizing the finite-difference approximation technique is shown in Figure~\ref{FD}. The steps for evaluating the controller are given as follows.

\begin{enumerate}
    \item Compute $V_k(u_k+\epsilon)$ using the expression:
    \begin{align}
    V_k(u_k+\epsilon) &= \bra{\psi_k^{+\epsilon}}H_0\ket{\psi_k^{+\epsilon}}  + \sum_{j=0}^{m-1} \alpha_j \abs{\braket{q_j}{\psi_k^{+\epsilon}}}^2,
    \label{V_dis}
    \end{align}
where $\ket{\psi_k^{+\epsilon}}=U_k^{+\epsilon}\ket{0}$ and $U_k^{+\epsilon}=U_1(u_k+\epsilon)U_0 \prod_{l=1}^{k-1} (U_1(u_l)U_0)\ket{\psi_0}$ . 
Estimating the term $\bra{\psi_k^{+\epsilon}}H_0\ket{\psi_k^{+\epsilon}} = \sum_{q=1}^{N_0} c_q \bra{\psi_k^{+\epsilon}} O_q \ket{\psi_k^{+\epsilon}}$ involves the estimation of $N_0$ expectations. For the term $\sum_{j=0}^{m-1} \alpha_j \abs{\braket{\psi_k^{+\epsilon}}{q_j}}^2$, we need to evaluate the overlap between the evolved state and the lower energy eigenstates. While the Hadamard test \cite{aharonov2006polynomial} can be used for this purpose, a more efficient method, as suggested in \cite{havlivcek2019supervised}, involves expressing each overlap term as $ \abs{\braket{q_j}{\psi_k^{+\epsilon}}}^2 = \abs{\bra{0} U^{(j) \dagger} U_k^{+\epsilon} \ket{0}}^2$. Consequently, by preparing the state $U^{(j) \dagger} U_k^{+\epsilon} \ket{0}$, the overlap can be estimated by the fraction of all-zero bitstrings obtained from measuring this state in the computational basis. Figure~\ref{Ge} shows the quantum circuits used to estimate both terms.

\begin{figure}[H]
    \centering
    
    \begin{subfigure}{\columnwidth}
        \centering
        \includegraphics[width=0.7\linewidth]{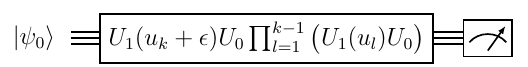}
        \caption{The quantum circuit used to estimate the term $\bra{\psi_k^{+\epsilon}}H_0\ket{\psi_k^{+\epsilon}}$.}
        \label{subfig:O1}
    \end{subfigure}
    
    \begin{subfigure}{\columnwidth}
        \centering
        \includegraphics[width=0.7\linewidth]{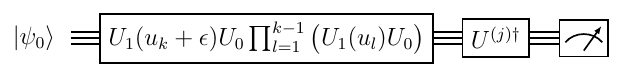}
        \caption{The quantum circuit used to estimate the term $\abs{\braket{q_j}{\psi_k^{+\epsilon}}}^2$.}
        \label{subfig:O2}
    \end{subfigure}

    \caption{The quantum circuits used to estimate the different terms of the controller $u_{k+1}$.}
    \label{Ge}
\end{figure}

\item Compute $V_k(u_k-\epsilon)$ using similar procedures to the above by replacing the state $\ket{\psi_k^{+\epsilon}}$ with the state  $\ket{\psi_k^{-\epsilon}}=U_1(u_k-\epsilon)U_0 \prod_{l=1}^{k-1} U_1(u_l)U_0\ket{\psi_0}=U_k^{-\epsilon}\ket{0}$.
\item Compute the controller using Equations \eqref{PB:u_dv} and \eqref{PB:FDa} .
\end{enumerate}

The gradient can also be computed using PSR, which gives an analytical formula for evaluating the gradient on a quantum computer \cite{mari2021estimating}. For the case of the control Hamiltonian $H_1$ having only two distinct eigenvalues, PSR is given as
\begin{equation}
      \frac{\partial}{\partial u_k} V_k(u_k) =\frac{1}{2} \big(V\left({u_k}+\pi/2 \right)-V\left({u_k}-\pi/2 \right) \big).
\end{equation}

\begin{figure}[H]
    \centering
    \includegraphics[ width=1 \linewidth]{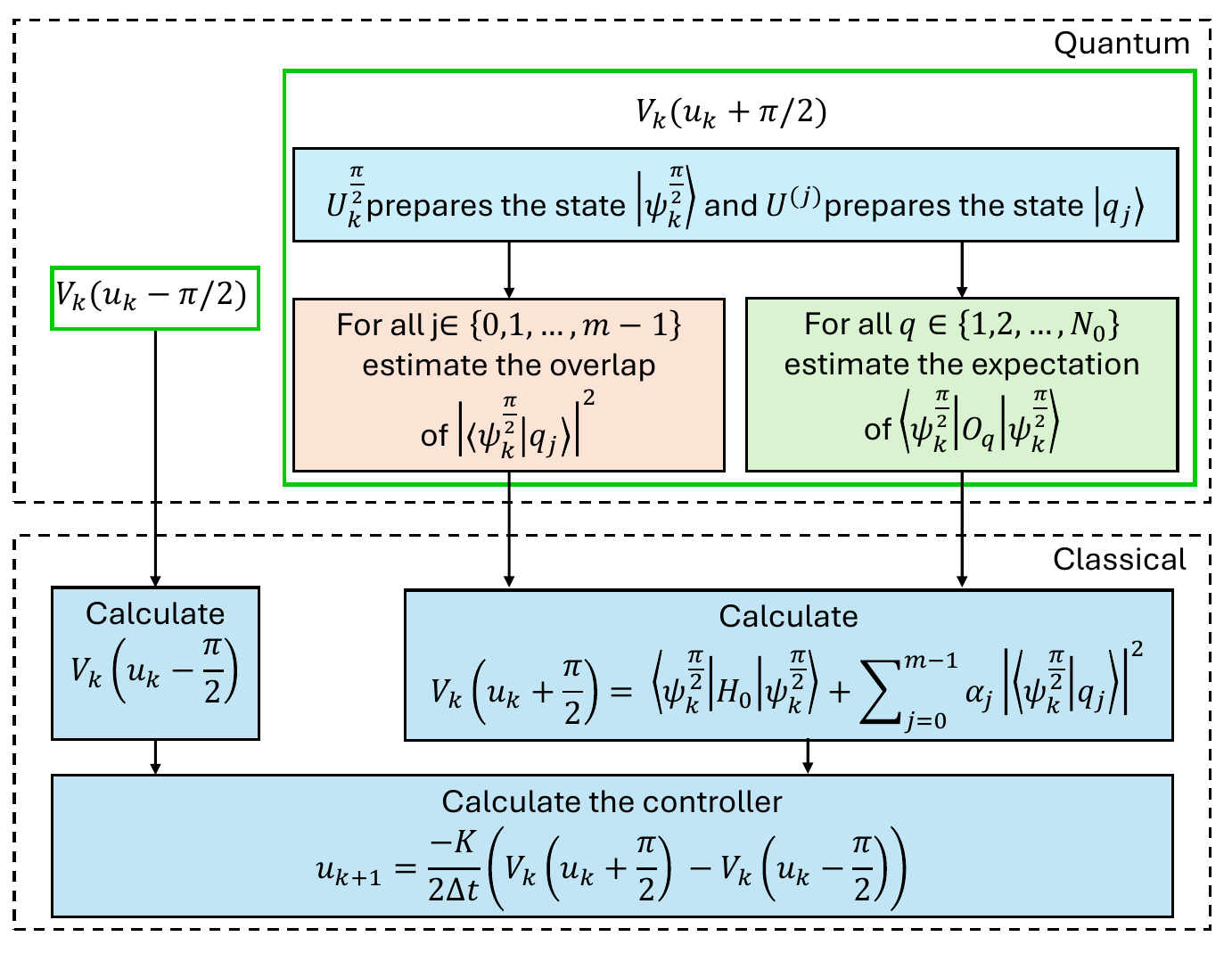}
    \caption{The gradient-based approach for calculating the controller using PSR to evaluate the gradient.}
    \label{PSR}
\end{figure}

In \cite{wierichs2022general}, PSR is generalized to a more general type of Hamiltonian. However, it will need more computations. A schematic diagram of the gradient-based approach utilizing PSR is shown in Figure~\ref{PSR}.

Estimating the gradient using the finite difference approximation can encounter numerical instabilities and convergence challenges \cite{baydin2018automatic}. Furthermore, in \cite{harrow2021low}, it has been demonstrated that using the analytical gradient outperforms any finite difference method. Therefore, a better solution can be using PSR. However, as shown in \cite{wierichs2022general}, PSR depends on the control Hamiltonian $H_1$, and for a more general Hamiltonian, more computation cost is needed to evaluate the gradient. In contrast, the finite difference formula given in \eqref{PB:FDa} is independent of the specific Hamiltonian $H_1$.

Both methods, overlap-based evaluation and gradient-based estimation, generally lead to deep circuits and may not be suitable for NISQ hardware. The overlap-based method requires implementing a Hadamard test, which involves controlled versions of the unitary circuits. This increases circuit depth and resource demands when executed on quantum hardware. In contrast, the gradient-based approach avoids the need for controlled unitaries and ancilla qubits, making it more suitable for execution on quantum hardware. Moreover, as demonstrated in the next section, the gradient-based scheme that uses the parameter-shift rule exhibits greater robustness to sampling noise. Hence, gradient-based approaches are generally more suitable for calculating the controllers.
\\
\\
\\
\\
\\
\\
\textbf{Remark 5.} As pointed out by an anonymous reviewer, for the case where the problem Hamiltonian $H_0$ is diagonal in the computational basis, each of the lower energy eigenstates is a computational basis state. In this case, the controller can be evaluated as follows. Assume for the simplicity of the analysis that we want to prepare the first excited state, then we have $P=H_0+\alpha_0 M_0$, where $M_0=\ketbra{q_0}$ and the controller is given as:
\begin{align}
     u_{k+1} & = -K   \bra{\psi_k}  \mathrm{i} [H_1,P]  \ket{\psi_k}, \nonumber \\
            & = -K    \bra{\psi_k} \mathrm{i}  [H_1,H_0+ \alpha_0 M_0]  \ket{\psi_k}, \nonumber \\
            & = -K    \Bigl( \bra{\psi_k}  \mathrm{i}[H_1,H_0]  \ket{\psi_k} +  \bra{\psi_k}  \mathrm{i} [H_1, \alpha_0 M_0]  \ket{\psi_k} \Bigr).
\end{align}
The first expectation $ \bra{\psi_k}  \mathrm{i}[H_1,H_0]  \ket{\psi_k}$ can be calculated as presented in Section 2. The second expectation can be estimated as follows. We represent $\ket{q_0} = \ket{b_0 \dots b_{n-1}}$, where $b_0 \in \{0,1\}$. Then we have $M_0 = \bigotimes_{l} \frac{I_l + (-1)^{b_l} Z_l}{2}$, where $\bigotimes$ represents tensor product. Assuming the control Hamiltonian is designed as the standard mixer Hamiltonian $H_1=\sum_{j=1}^n X_j$, then the second term simplifies to:
\begin{align}
    & \bra{\psi_k}i[H_1, \alpha_0 M_0] \ket{\psi_k}   =  \alpha_0 \bra{\psi_k} \sum_j \Big(  (-1)^{b_j+1} \big( \bigotimes_{l \neq j} P_l \big) \otimes Y_j \Big)\ket{\psi_k},
\end{align}
where $P_l = \frac{I_l + (-1)^{b_l} Z_l}{2}$, and we used $X_jZ_j=iY_j$ and $i^2=-1$. Therefore, we can estimate each of the expectations by measuring all the qubits in the computational basis except for the j'th qubit, which is measured in the Y basis. While this approach avoids the necessity of estimating the gradient of the Lyapunov function or using Hadamard tests, it cannot be applied to a general problem Hamiltonian $H_0$ where the lower energy eigenstates will not necessarily be computational basis states.

\section{Applications}
\label{sec:paperB:Applications}

To demonstrate the performance of FQAE, we give an illustrative example and apply the algorithm to a quantum chemistry problem where we determine the spectrum of the Hamiltonian of the hydrogen molecule. 
\subsection{An illustrative example}
Consider an Ising Hamiltonian in the form
\begin{equation}
    H_0=\sum_{q<j}^n J_{q, j} Z_q Z_j+\sum_{q=1}^n J_q Z_q,
\end{equation}
where $n$ is the number of qubits, $J_{q,j}$ and $J_q$ are real parameters with $J_{q,j}=J_{j,q}$, $Z_q$ is the Pauli Z operator applied to the $q$'th qubit.

For our analysis, we set $n=2$, $J_{1,2}=0.5$, $J_1=1$, $J_2=2$. The resulting Hamiltonian is $H_0= Z_1+2Z_2+0.5Z_1Z_2 = \text{diag}(3.5,-1.5,0.5,-2.5)$. We apply FQAE to prepare the first excited state of this Hamiltonian, where the ground state of this Hamiltonian is the state $\ket{q_0}=\ket{11}$ and its energy is $-2.5$. 

\begin{figure}[H]
    \centering
    
    \begin{subfigure}{0.45\linewidth} 
        \centering
        \includegraphics[width=0.65\linewidth]{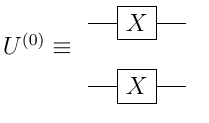}
        \caption{The quantum circuit to prepare the ground state $\ket{11}=U^{(0)} \ket{00}$.}
        \label{11}
    \end{subfigure}
    \hspace{0.05\linewidth} 
    \begin{subfigure}{0.45\linewidth} 
        \centering
        \includegraphics[width=0.6\linewidth]{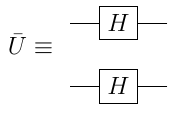}
        \caption{The quantum circuit to prepare the initial state $\ket{\psi_0}=\Bar{U} \ket{00}$.}
        \label{22}
    \end{subfigure}

    \vspace{1em} 

    \begin{subfigure}{0.9\linewidth} 
        \centering
        \includegraphics[width=0.55\linewidth]{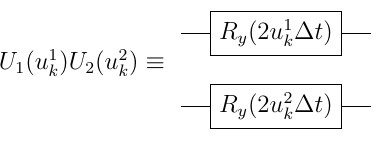}
        \caption{The quantum circuit to implement the control operator $U_c=U_1(u_k^1)U_2(u_k^2)$.}
        \label{44}
    \end{subfigure}
    \hspace{0.05\linewidth} 
    \begin{subfigure}{0.9\linewidth} 
        \centering
        \includegraphics[width=0.75\linewidth]{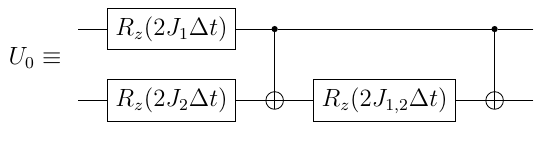}
        \caption{The quantum circuit to implement the unitary $U_0$.}
        \label{33}
    \end{subfigure}

    \caption{This figure shows the quantum circuits of the different unitary operators needed to implement the different steps of FQAE.}
    \label{circuits}
\end{figure}

The control Hamiltonian is designed as $H_c= \sum_{q=1}^2 u^{(q)}H_q =  u^{(1)}H_1+u^{(2)}H_2 = u^{(1)}Y_1+u^{(2)}Y_2$. The initial state is chosen to be the equal superposition state $\ket{\psi_0} = H^{\otimes 2} \ket{0}= \ket{++}$. The quantum circuits representing the operators $U_0$, $U_1(u^{(1)}_k)$, $U_2(u_k^{(2)})$, $\Bar{U}$ and $U^{(0)}$ are shown in Figure~\ref{circuits}.
\newcommand*{\figuretitle}[1]{%
    {\centering
    \textbf{#1}
    \par\medskip}
}

\begin{figure}[H]
    \centering
    \includegraphics[ width=1 \linewidth]{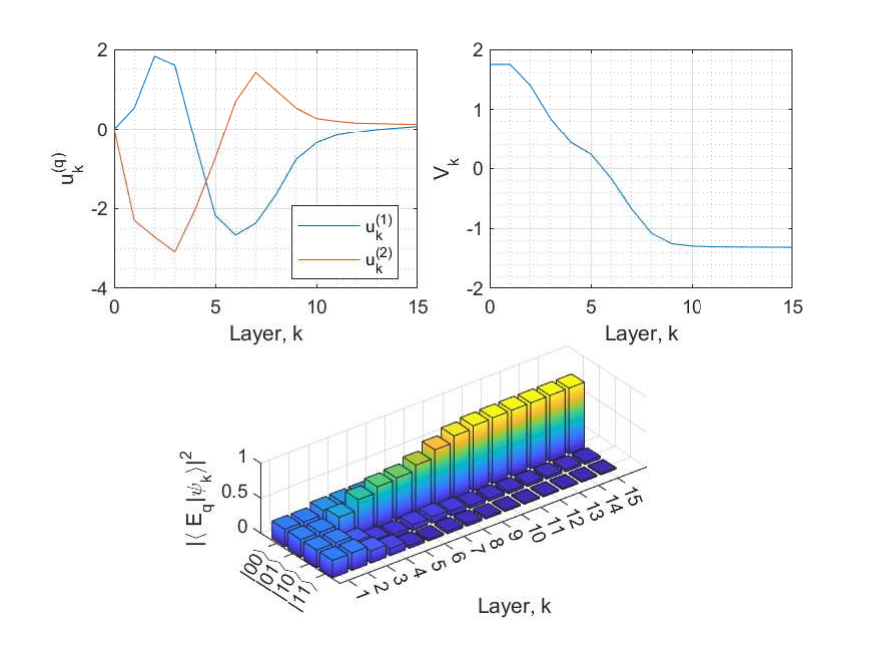}
    \caption{Simulation results of FQAE applied to prepare the first excited state $\ket{q_1}=\ket{01}$ of the Hamiltonian $H_0$. The layer $k$ is plotted versus: (a) The control inputs, (b) The Lyapunov function of the evolved state $V_k= \bra{\psi_k} P \ket{\psi_k}$, and (c) The fidelity of the evolved state with the eigenstates of the Hamiltonian $H_0$. }
    \label{SR1}
\end{figure}

To determine the value of $\alpha_0$, we employ the first approach, calculating the upper bound on the energy gap via equation \eqref{G}, yielding $G \leq 7$, and subsequently assigning $\alpha_0$ as $7$.
The operator $P$ is constructed as $P=H_0 + \alpha_0 \ketbra{11}$ and Hence, we get $P=\text{diag}(3.5,-1.5,0.5,4.5)$. Note that the ground state of the operator $P$ is the state $\ket{q_1}=\ket{01}$ which is the first excited state of the Hamiltonian $H_0$. The time step $\Delta t$ is chosen to be $0.08$, and the controllers' gains are chosen to be $K_1 = K_2 =1.5$. To execute the quantum circuit, the Qiskit Aer quantum simulator \cite{aleksandrowicz2019qiskit} is used. The simulation results for preparing the first excited state for the Hamiltonian $H_0$ using the exact statevector simulator are shown in Figure~\ref{SR1}.

Figure~\ref{SR1} shows that FQAE can calculate the first excited state. The Lyapunov function exhibits a monotonic behaviour, which aligns with our controller design in \eqref{PB:controller}. It is also seen that the fidelity with respect to the first excited state $\abs{\bra{\psi_k} \ket{q_1}}^2$ increases with the increase in the circuit layers approaching 1. Note that, in practice, evaluating the Lyapunov function at every step is not required; only the controller values are needed to construct the next layer. A convenient convergence check is therefore to monitor these controller values: when they approach zero, the algorithm is effectively converging toward the targeted eigenstate, and this behavior can be used to set the maximum circuit depth. As an additional (though more costly) option, one may track the decrement of the Lyapunov function itself and terminate when the change falls below a small threshold $\delta$, that is, when $V_{k+1} - V_k < \delta$ for some chosen $\delta > 0$. Because this alternative requires explicitly evaluating $V_k$ during runtime, introducing extra measurements and classical post-processing, we retain controller convergence as the most efficient and practical indicator within the present algorithmic framework.

In practice, the controllers will be evaluated through a finite number of samples. To assess the impact of sampling noise on FQAE's performance, we execute the algorithm for the two controller evaluation approaches proposed in Section~4 and for a different number of sample sizes. The results are compared to the exact values using statevector simulator(i.e. $m \rightarrow \infty$).

\begin{figure}[H]
    \centering
    \includegraphics[ width=1 \linewidth]{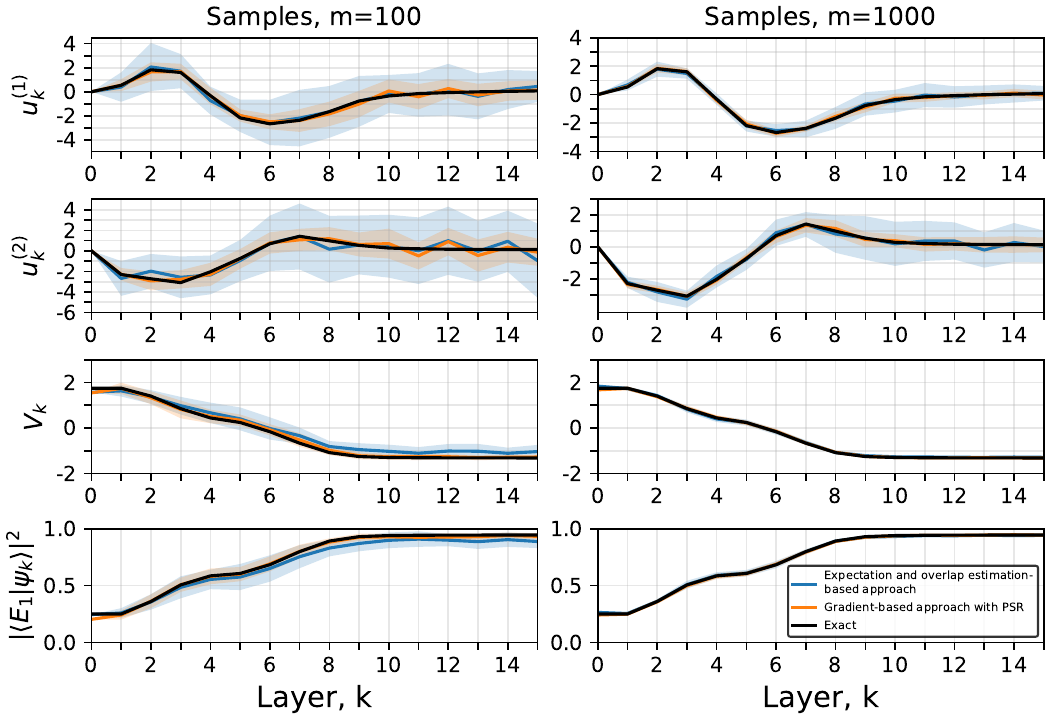}
    \caption{Simulation results of FQAE applied to prepare the first excited state of the Hamiltonian $H_0$ for a different number of samples and using the two proposed approaches to evaluate the controller: the expectation and overlap estimation-based approach and the gradient-based approach utilizing PSR. The layer $k$ is plotted versus the mean trajectory (solid line) and the corresponding standard deviation (shaded area) of the controllers $u_k^{(1)}$ and $u_k^{(2)}$ and the Lyapunov function $V_k=\bra{\psi_k}P\ket{\psi_k}$. In addition, the exact noiseless simulation ($m \rightarrow \infty$) is plotted for comparison (black solid line).}
    \label{u1u2}
\end{figure}

Figure~\ref{u1u2} shows that even with a small number of samples $m=100$, the estimated controller follows the exact controller, indicating that FQAE is robust against sampling noise. It also shows that the gradient-based approach for evaluating the controller using PSR has better robustness against the sampling noise than the expectation and overlap estimation-based approach.

\begin{figure}[H]
    \centering
    \includegraphics[ width=1 \linewidth]{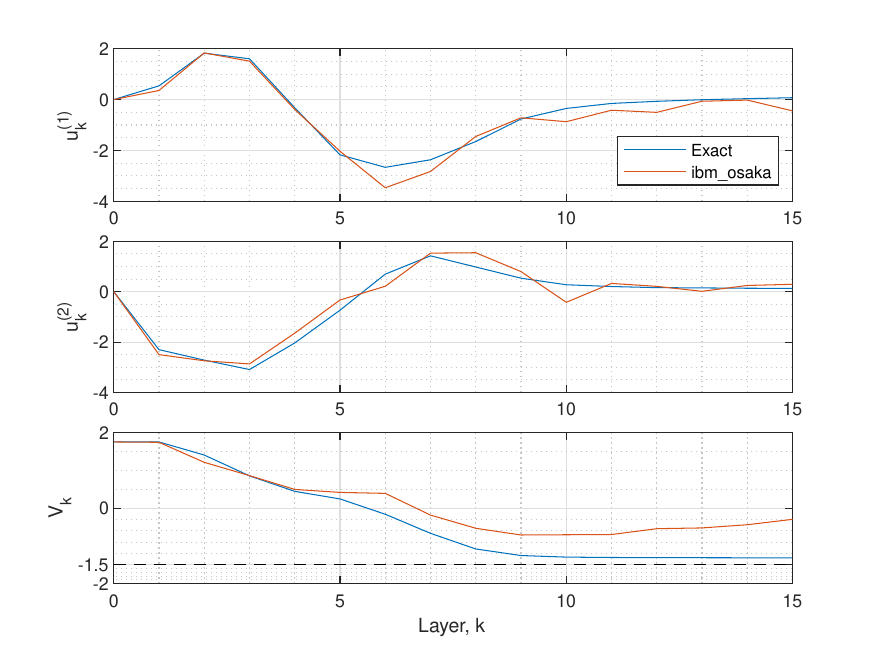}
    \caption{ Experimental results of running FQAE on the IBM superconducting quantum computer \textit{ibm\_osaka}. The gradient-based approach for evaluating the controller is used for this experimental run, where the PSR is used to evaluate the controller. The number of shots is set to $4000$.  The layer k is plotted versus: (a) The first control input $u_k^{(1)}$, (b) The second control input $u_k^{(2)}$, and (c) The Lyapunov function $V_k$.} 
    \label{experiemental}
\end{figure}

We now present, as a proof of principle, an experimental demonstration of FQAE on superconducting quantum computers. We run FQAE on the publicly available IBM quantum computer \textit{ibm\_osaka} through the cloud-based service \cite{IBM}. The gradient-based approach using PSR is adopted to evaluate the controller, and the number of shots is set to $4000$. The results are presented in Figure~\ref{experiemental}.

From Figure~\ref{experiemental}, we see that the controller's design is robust against noise and can recover the exact controllers calculated through the exact statevector simulator. We also see that the Lyapunov function monotonically decreases toward its global minimum value, which is $\bra{01} P \ket{01} = -1.5$ till the 10th layer. These experimental results underscore the promise of FQAE's practical applicability in the near future.

\begin{figure}[H]
    \centering
    \includegraphics[ width=1 \linewidth]{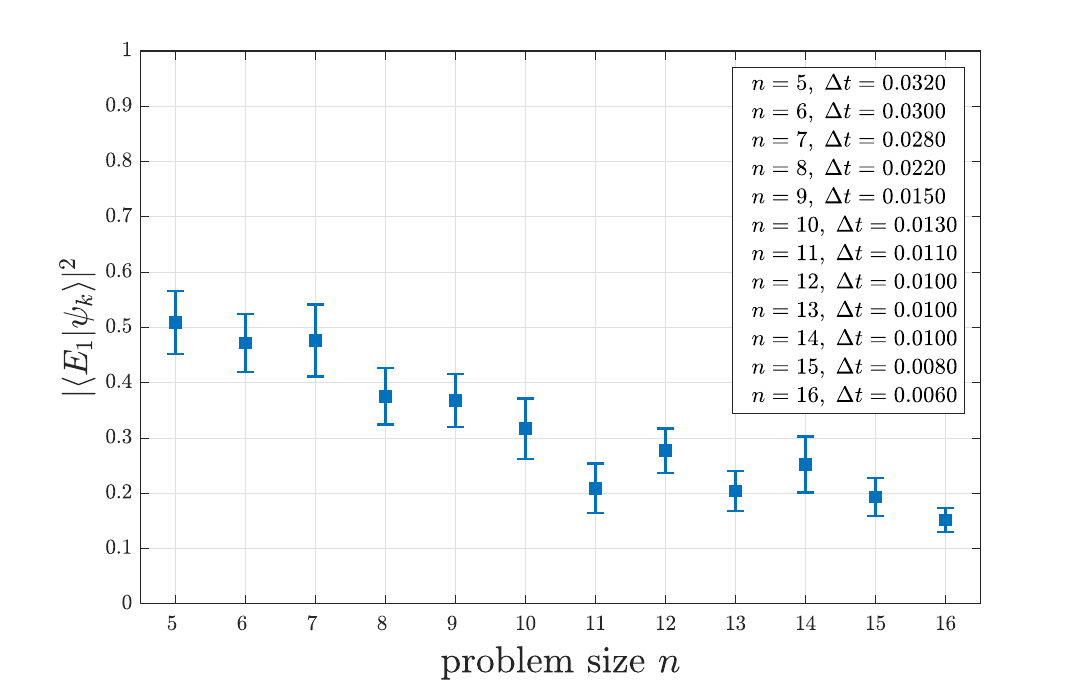}
    \caption{Average final fidelity as a function of problem size $n$. Each data point is the mean over $15$ random Ising instances; error bars denote the standard error of that mean.}
    \label{scalability}
\end{figure}

 We conclude this subsection by investigating how FQAE scales with system size.  For each size $n \in \{5,6,\ldots,16\}$ we generate fifteen random Ising Hamiltonians by drawing $J_{q,j}$ and $J_q$ independently from a uniform distribution over the interval $[-2,2]$. We run FQAE for all these random instances for a depth of $500$ layers. We set the shift parameter to $\alpha=4$. To tune the parameters, we follow a similar approach as proposed in \cite{magann2022lyapunov} where we fix the controller gain to $K=1$, and increase the time step $\Delta t$ to the largest possible value that guarantees the Lyapunov condition $V_{k+1}-V_k<0$ holds over all the random instances. The average final fidelity with respect to the first excited state, with error bars showing the standard error of the mean, is shown in Figure~\ref{scalability}.
Figure~\ref{scalability} shows that convergence slows as the system size increases, since the time step $\Delta t$ must be decreased to satisfy the Lyapunov condition $V_{k+1}-V_k < 0$.

\begin{figure}[H]
    \centering
    \includegraphics[ width=0.85 \linewidth]{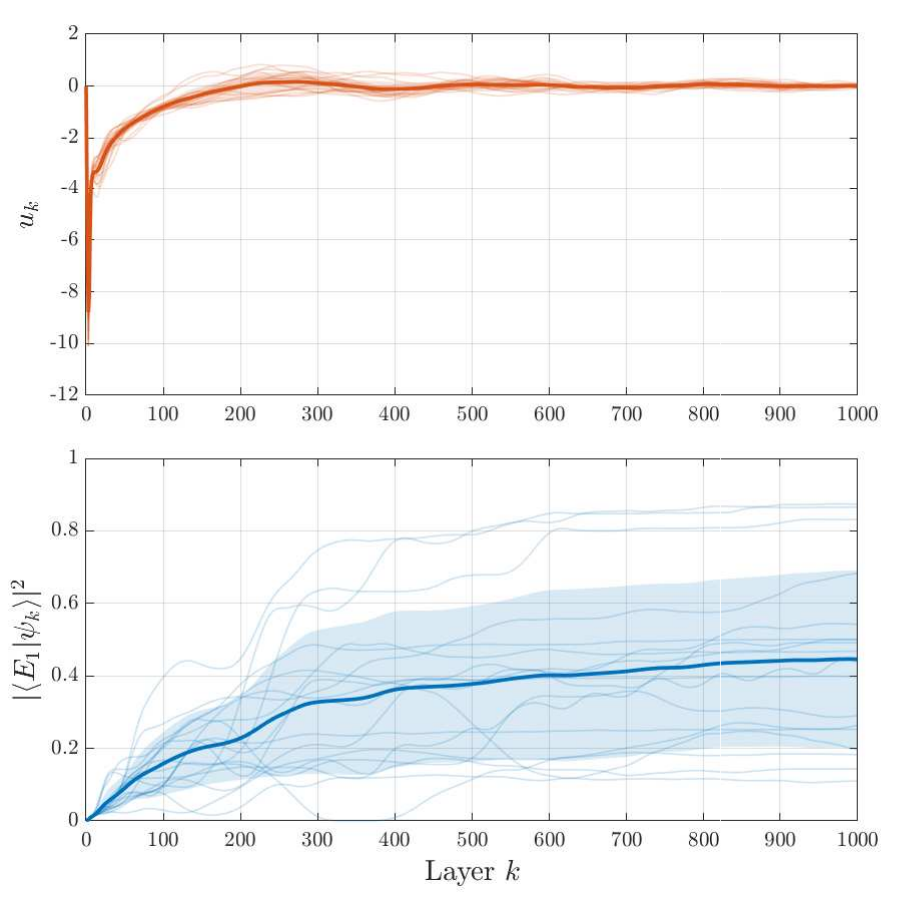}
       \caption{Performance over $20$ random Ising instances ($n=9$). The Light curves represent individual trajectories of the controller $u_k$ (top) and the fidelity $\abs{\braket{E_1}{\psi_k}}^2$ (bottom). The dark curves represent the mean, and the shaded area represents their corresponding standard deviations.}
    \label{premat}
\end{figure}

The numerical results show similar behaviour to the application of FALQON on weighted graphs of MAXCUT problem (see Appendix of \cite{magann2022feedback} for details), where for some random instances, the controller approaches zero prematurely prior to $\abs{\braket{E_1}{\psi_k}}^2 \rightarrow 1$.  Figure~\ref{premat} summarizes $20$ random Ising instances with $n=9$. As shown in the figure, while most of the random instances achieve fidelities above $0.4$, some instances converge to lower values. 

\begin{figure}[H]
    \centering 
    \includegraphics[ width=0.78 \linewidth]{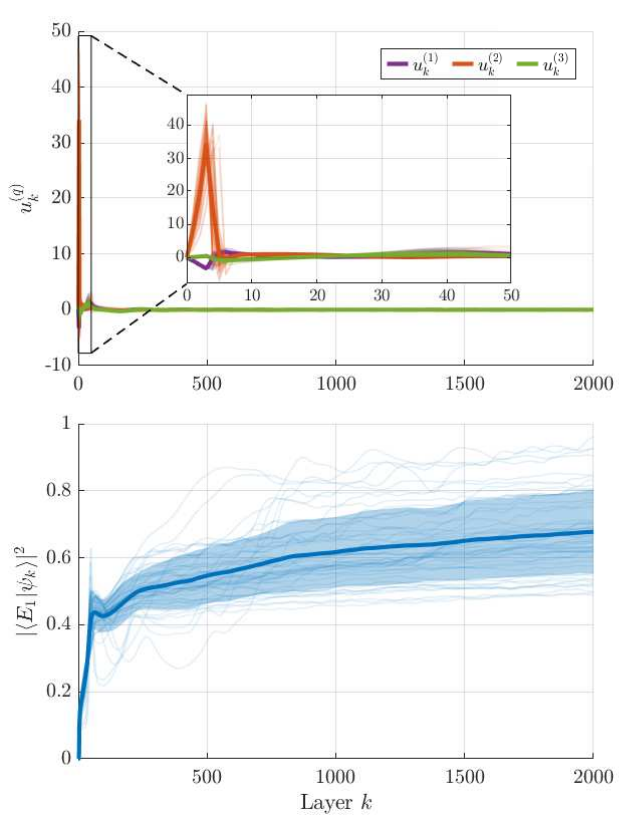}
       \caption{Performance over $50$ random mixed-field Ising instances ($n=12$). The light curves represent individual trajectories of the controllers $u^{(q)}_k$ (top) and the fidelity $\abs{\braket{E_1}{\psi_k}}^2$ (bottom). The dark curves represent the mean, and the shaded area represents their corresponding standard deviations.}
      \label{good}
\end{figure}

We additionally apply FQAE to the mixed-field Ising (MFI) model with periodic boundary conditions using multiple control inputs. The results demonstrate better convergence of the fidelities compared to the inhomogeneous Ising model with nine qubits, although the achieved fidelities remain lower than those obtained for the two-qubit case. The MFI model is defined as 
\begin{equation}
    H_0 = J \sum_{q}^{n} Z_qZ_{q+1} + h \sum_{q}^{n} X_q + g \sum_{q}^{n} Z_q,
\end{equation}
where $J$ represents the nearest-neighbor interaction strength, $h$ is the transverse field strength, and $g$ is the longitudinal field strength. We fix the system size to $n=12$ and generate $50$ random instances by setting $J=-1$ and independently sampling the parameters $h$ and $g$ from uniform distributions, $h \sim \mathcal{U}(0.4,1)$ and $g \sim \mathcal{U}(0.1,0.6)$. We run FQAE for all these random instances for a depth of $2000$ layers. The control Hamiltonian is designed as $H_c = \sum_{q=1}^3 u^{(q)}H_q = u^{(1)}H_1+u^{(2)}H_2+u^{(3)}H_3 = u^{(1)} \sum_{j=1}^n X_j+u^{(2)}\sum_{j=1}^n Y_j+u^{(3)}\sum_{j=1}^n Z_j$. The initial state is chosen as $\ket{\psi_0}=\ket{+}^{\otimes 12}$ and the controllers are initialized as $[u_0^{(1)}, u_0^{(2)}, u_0^{(3)}] = [0, 0, 0]$. The gains are set to $K_1=K_2=K_3=1$, the shift parameter to $\alpha=7$ and the time step to $\Delta t = 0.01$. The simulation results are presented in Figure~\ref{good}.
Unlike the behavior observed for the inhomogeneous Ising model, where some instances converged to lower fidelities and became trapped in local minima, FQAE with multiple control inputs achieved improved fidelities for the MFI model. This improvement may be partly attributed to the more favorable energy landscape of the MFI model and to the use of multiple controllers, which can enhance the performance of feedback-based quantum algorithms \cite{magann2022lyapunov}.

Although the mixed-field Ising results demonstrate improved performance relative to the inhomogeneous Ising case, the average fidelity to the first excited state for $n=12$ remains moderate (around 0.67) and does not reach the values observed for the small two-qubit example in Figures~\ref{SR1} and \ref{u1u2}. Moreover, increasing the circuit depth from $1000$ to $2000$ layers leads only to marginal improvements. These observations suggest that, in its current implementation, FQAE encounters significant limitations when applied to larger systems. In particular, the required circuit depth increases with system size, and the underlying energy landscape may still hinder convergence for certain instances.

In the literature, several methods are proposed to address these challenges. Magann et al.~\cite{magann2022lyapunov} introduce heuristic strategies to increase convergence speed, such as incorporating multiple control inputs, while other works \cite{brady2025feedback,rattighieri2025accelerating} explore additional techniques for accelerating feedback-based quantum algorithms. It has also been demonstrated in \cite{arai2025scalable} that including higher-order terms in the control law enables larger time steps $\Delta t$ without compromising convergence guarantees, thereby significantly increasing convergence speed. To address the landscape challenge, one promising direction is the randomized approach \cite{magann2023randomized}, which adaptively constructs the quantum circuit to avoid becoming trapped in a local minimum. Another approach is the adaptive construction of the quantum circuit proposed in \cite{tang2025nonvariational}, which helps improve convergence. Integrating these enhancements into FQAE is a promising direction for improving both its convergence speed and its robustness across diverse problem instances.

\subsection{Application to Molecular Hamiltonians}
Using transformations such as the Bravyi-Kitaev or the Jordan-Wigner (JW) transformations, the quantum simulation of fermionic systems can be translated into qubit operations where the Hamiltonian is generally expressed as a sum of Pauli strings as follows (see \cite{helgaker2013molecular} for details).
\begin{equation}
    H_M = \sum_q ^{N_m} h_q \hat{O}_q,
\end{equation}
where $h_j$ are real scalar coefficients and $N_m$ is the number of Pauli strings.

As an example, we chose the hydrogen molecule. The fermionic Hamiltonian of hydrogen, modelled using the minimal STO-3G basis and varying inter-nuclear separations, can be converted into qubit representation through the Bravyi-Kitaev transformation. This yields a two-qubit Hamiltonian dependent on the bond length parameter, as follows (see \cite{colless2018computation} for details).
\begin{align}
    H_0(R) &= h_0(R) I + h_1(R)Z_1 + h_2(R)Z_2 + h_3(R)Z_1Z_2 +h_4(R)Y_1Y_2 + h_5(R)X_1X_2,
 \end{align}
where $R$ is the interatomic distance and the coefficients $h_i(R)$ are real valued functions of the interatomic distance.

For numerical simulations, we use the values of the coefficients given in Table 1 in the supplementary information of \cite{colless2018computation}. The numerical values of the coefficients for the distance value $R=1.05$ are given as $h_0=-0.5626$, $h_1=-0.248783$,  $h_2=-0.248783$, $h_3=0.00850998$, $h_4=0$ and $h_5=0.199984$.

\begin{figure}[H]
    \centering
    \includegraphics[ width=1 \linewidth]{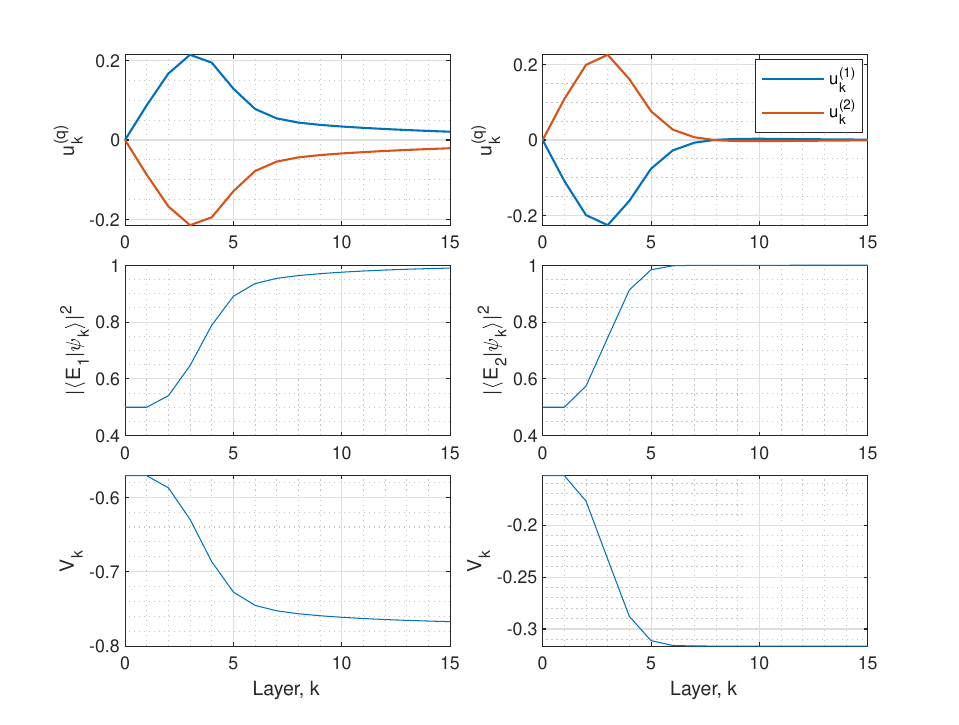}
    \caption{Illustration of FQAE to prepare the first excited state $\ket{E_1}$ (to the left) and the second excited state $\ket{E_2}$ (to the right). The layer $k$ is plotted versus: (a) The control inputs, (b) The Lyapunov function of the evolved state and (c) The fidelity with the targeted eigenstate.}
    \label{H2}
\end{figure}

We find the exact eigenstates and their corresponding energies of the hydrogen molecule Hamiltonian using exact diagonalization. We run two different simulations of FQAE to prepare the first and second excited eigenstates of $H_0$ using the statevector simulator. We choose the shifting parameters $\alpha_1 = 1.8$, $\alpha_2 =0.9$, the time step $\Delta t=0.55$, $K_1=K_2=1$ and the initial state to be $\ket{\psi_0}=\ket{01}$. The control Hamiltonian is designed as $H_c = \sum_{q=1}^2 u^{(q)}H_q = u^{(1)}H_1+u^{(2)}H_2 = u^{(1)} Z_1+u^{(2)}Z_2$. The initial guess for the controllers is chosen to be $[u^{(1)}_0,u^{(2)}_0]=[0,0]$. The simulation results are shown in Figure~\ref{H2}.

It is seen in Figure~\ref{H2} that FQAE can efficiently calculate the first and second excited states of the Hamiltonian $H_0$. The Lyapunov function monotonically decreases over circuit layers and converges to the operator $ P$'s ground state, which is the targeted eigenstate of $H_0$.

\begin{figure}[H]
    \centering
    \includegraphics[ width=1 \linewidth]{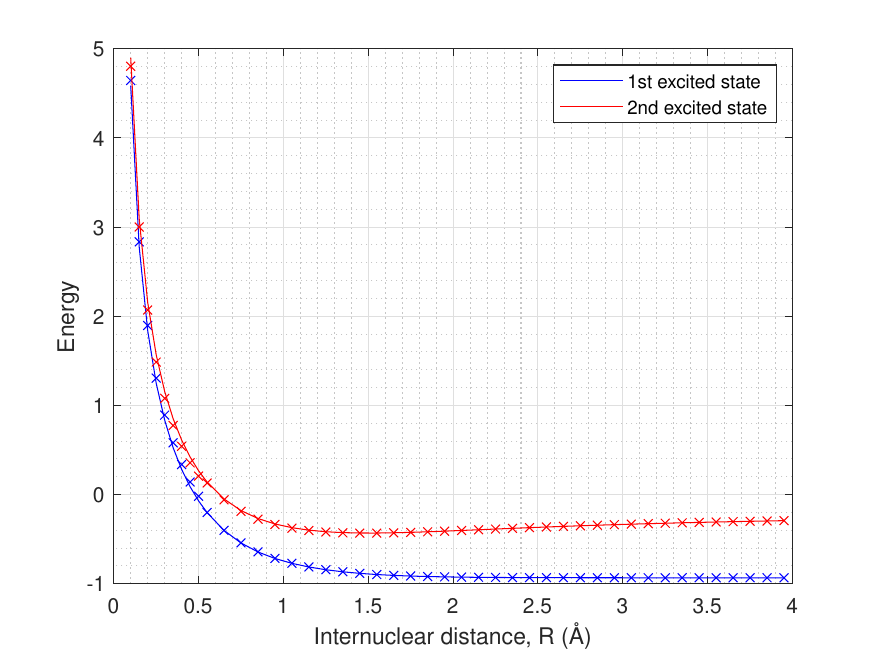}
    \caption{Demonstration of FQAE to calculate the energy of the first excited state $E_1$ and the second excited state $E_2$ of different values of $R$. The internuclear distance $R$ is plotted versus the energy calculated using FQAE, with solid lines indicating the exact energy of the excited states.}
    \label{H2_all}
\end{figure}

In addition, we run FQAE to calculate the energy of the first and second excited states for a range of R values $R \in [0.1,3.95]$ and for a depth of 25 layers. In this case, we design the control Hamiltonian to be $H_c = \sum_{q=1}^4 u^{(q)}H_q = u^{(1)}H_1+u^{(2)}H_2+ u^{(3)}H_3 + u^{(4)}H_4 = u^{(1)} Y_1+u^{(2)}Y_2 + u^{(3)}Z_1 + u^{(4)}Z_2$. For the values of $R \in [0.1,3.95]$ we set $\Delta t=0.55$, while for $R \in [0.1,0.45]$ we set $\Delta t= 0.15$ since these values of $\Delta t$ ensure controllers' convergence as suggested by simulation results. Since the Hamiltonians $H_0$ and $H_c$ have non-commuting terms, we use the first-order Lie-Trotter Suzuki decomposition to implement the unitaries $U_0$ and $U_c$ as quantum circuits in our simulations. The simulation results are shown in Figure~\ref{H2_all}. Note that the ground and highest excited states can be calculated using FALQON.

\section{Conclusion and Future Work}
\label{sec:paperB:Conclusion and Future Work}

In this paper, we have proposed FQAE, which extends the feedback-based quantum algorithms for calculating excited states. A comprehensive investigation of the algorithm's efficient implementation and controller evaluation is provided, presenting two alternative approaches for the controller evaluation. Among these approaches, the gradient-based method utilizing PSR yielded the most promising results in terms of performance and computation cost, emerging as the preferred option for near-term applications. The effectiveness of FQAE is demonstrated through an illustrative example and an application in quantum chemistry. Performance evaluations, including numerical simulations and execution on IBM's superconducting quantum computer, confirm the algorithm's potential for practical quantum computations.

As highlighted in Section~2, a significant challenge of the feedback-based quantum algorithms lies in the stringent assumptions imposed on the Hamiltonians $H_0$ and $H_1$, making their practical fulfilment challenging. However, as shown in \cite{magann2022lyapunov, magann2022feedback,larsen2024feedback,magann2023randomized}, and through numerical simulations, asymptotic convergence towards the desired eigenstate and good performance remain achievable even without strict adherence to these assumptions. Additionally, to enhance algorithmic performance, a range of heuristics has been proposed in \cite{magann2022feedback}, offering pathways to achieve convergence towards the target eigenstate in scenarios where adherence to the assumptions is unattainable. Furthermore, the randomization method proposed in \cite{magann2023randomized} helps in convergence to the targeted eigenstate and avoids convergence to sub-optimal solutions. Future endeavours may involve relaxing some of these assumptions, perhaps by incorporating mid-circuit measurements, as explored in \cite{clausen2023measurement}.

Another main challenge of FQAE, which is the bottleneck of the algorithm for being applicable to near-term devices, is the deep circuit required to run the algorithm. One approach to tackle this problem could be, as highlighted in Remark~2, by adopting a warm starting technique for the initial state of the algorithm. Another approach to overcome this challenge could be adopting a fixed-time control strategy as presented in \cite{li2022lyapunov}. The fixed-time Lyapunov control technique could guarantee convergence to the excited state in a fixed time and independently of the initial state. This could help in designing fixed-depth quantum circuits. Therefore, valuable future work would be adapting FQAE to such control designs. In addition, a valuable future work will be to estimate the required circuit depth to achieve a specific fidelity with the target eigenstate.

We highlight that feedback-based quantum algorithms remain applicable in near-term devices as a warm-starting technique for VQAs. In \cite{magann2022lyapunov}, it was demonstrated that employing FALQON as an initialization technique for QAOA can potentially improve its performance. Our work demonstrates the potential of quantum control theory to inspire the development of efficient quantum algorithms.

	\begingroup
	\footnotesize
	\raggedright	
	\bibliography{litterature.bib}	
	\endgroup

\end{document}